\advance\topmargin by 2.0in
\documentclass[aps,pra,floatfix]{revtex4}

\usepackage{graphicx}
\usepackage{epsfig}

\begin{document}

%\usepackage{graphicx}
%\usepackage{psfig}
%\begin{document}
%\twocolumn[\hsize\textwidth\columnwidth\hsize\csname@twocolumnfalse%
%\endcsname

\title{Spin Exchange Interactions of Spin-One
Bosons in Optical Lattices: Singlet, Nematic and Dimerized Phases.
}
\author{Adilet Imambekov, Mikhail Lukin, and Eugene Demler}
\affiliation{Department of Physics, Harvard University, Cambridge MA 02138}

\date{\today}

%\pacs{PACS numbers:}
%]

\begin{abstract}

We consider insulating phases of cold spin-1 bosonic particles
with antiferromagnetic interactions, such as $^{23}Na$, in optical
lattices. We show that spin exchange interactions give rise to
several distinct phases, which differ in their spin correlations.
In two and three dimensional lattices, insulating phases with an
odd number of particles per site are always nematic. For
insulating states with an even number of particles per site, there
is always a spin singlet phase, and there may also be a first
order transition into the nematic phase. The nematic phase breaks
spin rotational symmetry but preserves time reversal symmetry, and
has gapless spin wave excitations. The spin singlet phase does not
break spin symmetry and has a gap to all excitations. In one
dimensional lattices, insulating phases with an odd number of
particles per site always have a regime where translational
symmetry is broken and the ground state is dimerized.  We discuss
signatures of various phases in Bragg scattering and time of
flight measurements.

\end{abstract}

\maketitle

\section{Introduction}

Modern studies of quantum magnetism in condensed matter physics go
beyond explaining details of particular experiments on the cuprate
superconductors, the heavy fermion materials, organic conductors,
or related materials, and aim to develop general paradigms for
understanding  complex orders in strongly interacting many body
systems
\cite{assa,affleck,nayak01,kivelson02,ivanov02,levin03,hermele03,nersesyan,moessner,balents01}.
Spinor atoms in optical lattices provide a novel realization of
quantum magnetic systems that have several advantages compared to
their condensed matter counterparts, including precise knowledge
of the underlying microscopic models, the possibility to control
parameters of the effective lattice Hamiltonians, and the absence
of disorder.

Degenerate alkali atoms are generally considered as a weakly
interacting gas due to the smallness of the scattering length
compared to the inter particle separation \cite{Pethick}. The
situation may change dramatically either when atomic scattering
length is changed by means of Feshbach resonance \cite{Burnett}, or when an
optical potential created by standing laser beams confines
particles in the minima of the periodic potential and strongly
enhances the effects of interactions. In the latter case existence
of the nontrivial Mott insulating state of atoms in optical
lattices, separated from the superfluid phase by the quantum phase
transition (SI transition), was demonstrated recently in
experiments \cite{Orzel,Bloch,Rolston}. Low energy (temperature)
properties of spinless bosonic atoms in a periodic optical
potential are well described by the Bose-Hubbard Hamiltonian
\cite{Jaksch98}
\begin{eqnarray}
{\cal H}_{BH}=-t\sum_{\langle i
j\rangle}(a^{\dagger}_{i}a_{j}+a^{\dagger}_{j}a_{i}) - \mu
\sum_{i}\hat n_i+ \frac{U_0}{2}\sum_{i}\hat n_i(\hat n_i-1),
\label{BHHamiltonian}
\end{eqnarray}
Parameters of  (\ref{BHHamiltonian}) may be controlled by varying
the intensity of laser beams, so one can go from the regime in
which the kinetic energy dominates (weak periodic potential,
$t>>U_0$), to the regime where the interaction energy is the most
important part of the Hamiltonian (strong periodic potential,
$t<<U_0$). For integer fillings (number of atoms per lattice
site), the two regimes have superfluid and Mott insulating ground
states, respectively, as can be obtained from the mean-field
analysis of the Bose-Hubbard Hamiltonian
\cite{Jaksch98,Fishers89}. In the superfluid phase, atoms are
delocalized in the lattice, fluctuations in the number of atoms in
each site are strong, and there is a phase coherence between
different sites. In the insulating state, atoms are localized,
fluctuations in the particle number at each site are suppressed,
and there is a gap to all excitations. Such an insulating state
represents a correlated many body state of bosons, where strong
interactions between atoms result in a new ground state of the
system.

%Another important milestone in the studies of ultracold atoms was
%the creation of Bose condensates in the systems trapped purely by
%optical means \cite{MITOpticalTrapping}.
 In conventional magnetic
traps, spins of atoms are frozen so effectively that they behave
like spinless particles. In contrast, optically trapped atoms have
extra spin degrees of freedom which can exhibit different types of
magnetic orderings. In particular, alkali atoms have a nuclear
spin $I=3/2$. Lower energy hyperfine manifold has 3 magnetic
sublevels and a total moment $S=1.$ Various properties of such
condensate in a single trap were investigated
\cite{Law98,OhmiMachida,Ho98,Castin00,HoYip,Liu}. For example, for
particles with antiferromagnetic interactions, such as $^{23}{\rm
Na}$, the exact ground state of an even number of particles in the
absence of a magnetic field is a spin singlet described by a
rather complicated correlated wave function \cite{Law98}. However,
when the number of particles in the trap is large, the energy gap
separating the singlet ground state from the higher energy excited
states is extremely small, and for the experiments of ref.
\cite{MITOpticalTrapping}, the precession time of the classical
mean-field ground state is of the order of the trap lifetime. So,
experimental observation of the quantum spin phenomena in such
systems is very difficult. To amplify  quantum spin effects one
would like to have a system with smaller number of particles and
stronger interactions between atoms. Hence it is natural to
consider an idea of $S=1$ atoms in an optical lattice, in which
one can have a small number of atoms per lattice site (in
experiments of Ref. \cite{Bloch} this number was around 1-3) and
relatively strong interactions between atoms.

In this paper we study bosonic $S=1$ atoms in optical lattices
with spin symmetric confining potentials and antiferromagnetic
interaction between atoms. We demonstrate that spin degrees of
freedom result in a rich phase diagram by establishing the
existence of several distinct insulating phases, which differ from
each other by their spin correlations.

 In the insulating state of
bosons in an optical lattice fluctuations in the particle number
on each site are suppressed but not frozen out completely. Virtual
tunnelling of atoms between neighboring lattice sites gives rise
to effective spin exchange interactions that determine the spin
structure of the insulating states (spin exchange interactions for
$S=1/2$ bosons in optical lattices were discussed previously in
\cite{Svistunov,Duan}).

\begin{figure}
\psfig{file=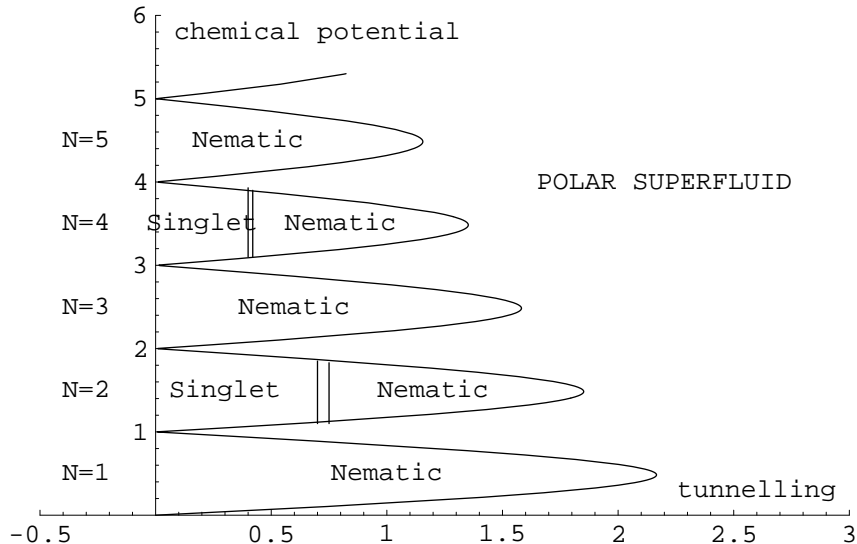} \caption{\label{phasediag0} General
phase diagram for $S=1$ bosons in 2d and 3d optical lattice.
Detailed discussion of the phase diagram, including explicit expressions
for various phase boundaries, is given in
Section \ref{GlobalPhase diagram}.}
\end{figure}

 We will show that in two and three dimensional
lattices insulating states with an odd number of atoms per site
are always nematic, whereas insulating states states at even
fillings are either singlet or spin nematic \cite{DemlerZhou},
depending on the parameters of the model. In one dimensional
systems even more exotic ground states should be realized,
including the possibility of a spin singlet dimerized phase that
breaks lattice translational symmetry \cite{zhou02,yip03}. The 2d
and 3d general phase diagram, including singlet, nematic and
 superfluid phases, is shown in Fig. \ref{phasediag0}. The extended
 version of this diagram, including discussion of various
 transition lines, is presented in section \ref{GlobalPhase
 diagram}.

It is useful to point out that the lattice model for spin-1
bosons, which we analyze here, is very general and may also be
applicable to systems other than cold atoms in optical lattices.
For example, triplet superconductors in strong coupling limit may
be described by a similar Hamiltonian, and some of the phases
discussed in this article may correspond to non-BCS states of such
superconductors \cite{gang5}.

The paper is organized as follows. In section
\ref{introtoformalism} we provide derivation of the Hubbard-type
Hamiltonian for spin-1 bosons in optical lattices starting from
microscopic interactions between atoms, and describe some general
properties of our model.  In section \ref{Nodd} we derive an
effective spin Hamiltonian which is valid for any odd number of
atoms per site, $N$, in the limit of small tunnelling between
sites. We demonstrate equivalence between our system and a
Heisenberg model for $S=1$ spins on a lattice with biquadratic
interactions and argue that the ground state is a nematic in two
and three dimensions and is a dimerized singlet in 1d.  In section
\ref{N2} we derive effective spin Hamiltonian for a system with
$N=2$ atoms per site, valid deep in the insulating regime, and use
mean-field approximation to determine the phase boundaries between
isotropic and nematic phases. In section \ref{Nbig} we derive
effective spin Hamiltonian for the limit of large number of
particles per site $N>>1$ and small tunnelling, and discuss
isotropic-nematic transition for even $N$. In section
\ref{GlobalPhase diagram}  we summarize our results and review the
global phase diagram for spin-1 bosons in optical lattices.
Finally, in section \ref{experiment}, we discuss approaches to
experimental detection of singlet and nematic insulating phases of
$S=1$ bosons. Details of technical calculations are presented in
Appendices A-D.

\section{Derivation of Bose-Hubbard model for spin-1 particles}
\label{introtoformalism}

At low energies scattering between two identical alkali atoms with
the hyperfine spins $S=1$ is well described by the contact
potential \cite{Pethick}
\begin{eqnarray}
 \hat V({\bf
r_1-r_2}) &=& \delta({\bf r_1-r_2})(g_0 {\cal P}_0 + g_2 {\cal P}_2 )
\label{Vr1r2}, \\
g_S &=& 4 \pi \hbar^2 a_S/M.
\end{eqnarray}
Here ${\cal P}_S$ is the projection operator for the pair of atoms
into the state with total spin $S=0,2$; $a_S$ is the $s$-wave
scattering length in the spin $S$ channel; and $M$ is the atomic
mass. When writing (\ref{Vr1r2}) we used the fact that $s$-wave
scattering of identical bosons in the channel with total spin $1$
is not allowed by the symmetry of the wave function. Interaction
(\ref{Vr1r2}) can be written using spin operators as
\begin{equation}
 V({\bf r_1-r_2})=\delta({\bf r_1-r_2})(\frac{g_0+ 2 g_2}3
+ \frac{g_2-g_0}3{ \bf S_1
 S_2}).
\end{equation}
For example, in the case of $^{23}Na$, $g_2>g_0,$ and  we find
effective antiferromagnetic interaction, as was originally
discussed in \cite{Ho98,OhmiMachida}.

Kinetic motion of ultracold atoms in the optical lattice is
constrained to the lowest Bloch band when temperature and
interactions are smaller than the band gap (this is the limit that
we will consider from now on). Atoms residing on the same lattice
site have identical orbital wave functions and their spin wave
functions must be symmetric. If we introduce creation operators,
$a^{\dagger}_{i\sigma}$, for states in the lowest Bloch band
localized on site $i$ and having spin components $\sigma=\{-1, 0,
1\}$, we can follow the approach of \cite{Jaksch98} and write the
effective lattice Hamiltonian
\begin{eqnarray}
{\cal H}=-t\sum_{\langle i j\rangle,\sigma}(a^{\dagger}_{i
\sigma}a_{j \sigma}+a^{\dagger}_{j \sigma}a_{i \sigma}) +
\frac{U_0}{2}\sum_{i}\hat n_i(\hat n_i-1) + \frac{U_2}{2}
\sum_{i}(\vec S_i^2-2 \hat n_i ) - \mu \sum_{i}\hat n_i,
\label{OriginalHamiltonian}
\end{eqnarray}
where
\begin{eqnarray}
n_i = \sum_\sigma a^{\dagger}_{i \sigma} a_{i \sigma}
\end{eqnarray}
is the total number of atoms on site $i$, and
\begin{eqnarray}
\vec{S}_i = \sum_{\sigma \sigma'} a^{\dagger}_{i \sigma}
\vec{T}_{\sigma \sigma'} a_{i \sigma'}
\end{eqnarray}
is the total spin on site $i$ ($\vec{T}_{\sigma \sigma'}$ are the
usual spin operators for spin 1 particles). The first term in
(\ref{OriginalHamiltonian}) describes spin symmetric tunnelling
between nearest-neighbor sites, the second term describes Hubbard
repulsion between atoms, and the third term penalizes non-zero
spin configurations on individual lattice sites. The origin of
this spin dependent term is the difference in scattering lengths
for $S=0$ and $S=2$ channels as was discussed in \cite{Law98}.
Finally, the fourth term in (\ref{OriginalHamiltonian}) is the
chemical potential that controls the number of particles in the
system.

Hamiltonian (\ref{OriginalHamiltonian}) carries important
constraints on possible spin states of the system. The first of
them derives from the fact that the total spin of a system of $N$
spin-1 atoms cannot be bigger than $N$, so for each lattice site
we have
\begin{eqnarray}
S_i \leq N_i.
\end{eqnarray}
The second constraint is imposed by the symmetry of the spin wave
function on each site
\begin{eqnarray}
S_i + N_i = {\rm even}.
\end{eqnarray}

\begin{figure}
\psfig{file=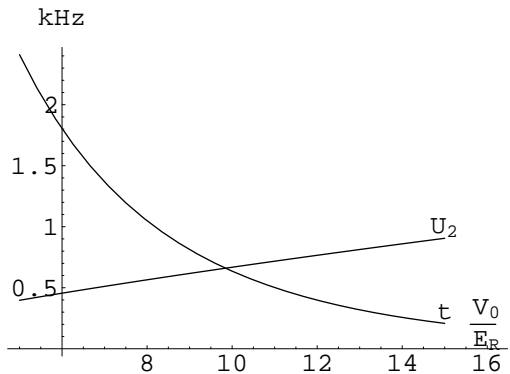} \caption{\label{tU2} $U_2$ and $t$ for
$^{23}Na$ atoms in the simple cubic optical lattice created by
three perpendicular standing laser beams with $\lambda=985 nm$.
$V_0$ is the strength of the optical potential and
$E_R=\hbar^2k^2/2M$ is the recoil energy. The ratio of the
interaction terms in (\ref{OriginalHamiltonian}), $U_2/U_0$, is
fixed by the ratio of the scattering lengths and is independent of
the nature of the lattice ($U_2/U_0 \approx 0.04 $ for $^{23}Na$)}
\end{figure}

Optical lattices produced by far detuned lasers with wavelength
$\lambda_i=2\pi/|\vec{k}_i|$ create an optical potential
$V(\vec{r})= \sum_i V_i \sin^2\,\,{\vec{k}_i \, \vec{r}},$ with
$\vec{k}_i$ being the wave vectors of laser beams. Using various
orientations of beams, one can construct different geometries of
the lattice. For the simple cubic lattice, parameters of
(\ref{OriginalHamiltonian}) can be estimated as
$$
U_2=\frac{2 \pi^2}3 E_R\frac{a_2 - a_0}{\lambda}x^{3/4},
$$
$$
U_0=\frac{2 \pi^2}3 E_R\frac{a_0 + 2 a_2}{\lambda}x^{3/4},
$$
$$
t=\frac{2}{\sqrt{\pi}}E_R x^{3/4}e^{-2 x^{1/2}},
$$
where $E_R=\hbar^2 k^2/2M$ is the recoil energy and $x=V_0/E_R.$
Note that the ratio $U_2/U_0$ is fixed by the ratio of scattering
lengths, $a_2/a_0$, for all lattice geometries. Scattering lengths
for $^{23}Na$ given in \cite{Nascatteringlengts} are $a_2=(52 \pm
5) a_B$ and $ a_0=(46 \pm 5 ) a_B,$ where $a_B$ is the Bohr
radius. This corresponds to $0<a_2-a_0<< 2 a_2 + a_0$, so the spin
dependent part of the interaction is much smaller than the spin
independent one. Throughout this paper we will always assume
$0<U_2<<U_0.$ While applying results of this paper for the case of
$^{23}Na$, one should note that errors in the estimation of the
exact value of $U_2/U_0$ are very big. While considering the spin
structure of Mott insulating phases, we will assume that $U_2/U_0$
is small enough to see the interplay between tunnelling and spin
dependent $U_2$ term before the superfluid-insulator transitions
take place. The positions of superfluid-insulator transitions and
the validity of this assumption will be discussed in detail in
section \ref{GlobalPhase diagram}.
 We will use the value $U_2/U_0=0.04$  to make estimates of various
phase boundaries.  In Fig. \ref{tU2} we show $U_2/\hbar$ and
$t/\hbar$ as a function of the strength of the optical potential
for a three dimensional cubic lattice produced by red detuned
lasers with $\lambda = 985 nm$.

Superfluid-insulator transition is characterized by a change in
fluctuations in particle number on individual lattice sites. When
the spin dependent interaction ($U_2$) is much smaller than the
usual Hubbard repulsion ($U_0$), the superfluid - insulator
transition is determined mostly by $U_0$.  The spin gap $U_2$
term, however, is important inside the insulating phase, where it
competes with the spin exchange interactions induced by small
fluctuations in the particle number, and an interesting spin
structure of the insulating states appears as a result of such
competition. The spin structure of the insulating phases of spin-1
bosons in optical lattices will be explored in this paper.

In what follows we will often find it convenient to use particle
creation operators that transform as vectors under spin rotations.
Such representation may be constructed as
\begin{equation}
a_z^{\dagger} = a_0^{\dagger},
a_x^{\dagger}=\frac{(a_{-})^{\dagger}-a_+^{\dagger}}{\sqrt2},a_y^{\dagger}=i
\frac{(a_{-})^{\dagger}+a_+^{\dagger}}{\sqrt2}. \label{X}
\end{equation}
Operators $a_{\{x,y,z\}}$ satisfy the usual bosonic commutation
relations, and they can be used to construct spin operators as
\begin{equation}
 S_{ia}= - i e_{abc} a^{\dagger}_{i b}a_{i
c}, \vec S_i^2=-(\delta_{bn}\delta_{\gamma
m}-\delta_{bm}\delta_{\gamma n})a^{\dagger}_b
a_{\gamma}a^{\dagger}_n a_m. \label{Spinviaa}
\end{equation}
We can verify the transformation properties of $a_{\{x,y,z\}}$ by
noting that
\begin{eqnarray}
[S_a,a_b]=[- i e_{apc} a^{\dagger}_{p}a_{c},a_b]=i e_{abc} a_{c},
\label{avector1}
\end{eqnarray}
\begin{eqnarray}
[S_a,a^{\dagger}_b]=[- i e_{apc}
a^{\dagger}_{p}a_{c},a^{\dagger}_b]=i e_{abc} a^{\dagger}_{c}.
\label{avector2}
\end{eqnarray}

Using these operators the hopping term  in the Hamiltonian
(\ref{OriginalHamiltonian}) may be rewritten as
$$-t\sum_{<i j>,p\in \{x,y,z\}}(a^{\dagger}_{i
p}a_{j p}+a^{\dagger}_{j p}a_{i p})$$ and it is invariant under
global spin rotations. We will use this property later to simplify
calculations and classify eigenstates of effective interaction by
the total spin of two neighboring sites.

\section{Insulating state with an odd number of atoms}
\label{Nodd}

\subsection{ Effective Spin Hamiltonian for small $t$}

We start with the insulating state of the Hamiltonian
(\ref{OriginalHamiltonian}) with an odd number $(N=2n+1)$ of
bosons per site in the limit $t=0$.  The number of particles on
each site is fixed, and the bosonic symmetry of the wave function
requires that the spin in each site is odd.  The interaction $U_2$
term is minimized when the spins take the smallest possible value
$S=1$. In this limit the energy of the system does not depend on
the spin orientations on different sites. When $t$ is finite but
small, we expect that we still have spin $S=1$ in each site, but
that boson tunnelling processes induce effective interactions
between these spins.  In this section we will compute such
interactions in the lowest (second) order in $t$. We will also
discuss conditions for which our effective Hamiltonian provides an
adequate description of the system.

In the second order perturbation theory in $t$, we generate only
pairwise interactions between atoms on neighboring sites, so we
can write the most general spin Hamiltonian for $S=1$ particles
that preserves spin $SO(3)$ symmetry
\begin{eqnarray}
{\cal H}=-J_0-J_1\sum_{\langle i j\rangle}\vec{S}_{i }\vec{S}_{j }-J_2\sum_{<i j>}(\vec{S}_{i
}\vec{S}_{j })^2.
\label{Hamiltonian_N1a}
\end{eqnarray}
Here $\langle ij \rangle$ labels near neighbor sites on the
lattice. Absence of the higher order terms, such as
$(\vec{S}_{i}\vec{S}_{j})^3,$ follows from the fact that the
product of any three spin operators for an $S=1$ particle can be
expressed via the lower order terms.

To find the exchange constants $J_{0,1,2}$, we need to consider
virtual processes that create a state with $N_i=2n, N_j=2n+2,$ and
$ N_i=2n+2, N_j=2n.$ The difference in energy between the
intermediate state and low energy $S_i=S_j=1$ subspace is of order
$U_0,$  and while our subspace is much lower in energy, the second
order perturbation theory is valid.

It is convenient to rewrite the Hamiltonian
(\ref{Hamiltonian_N1a}) as
\begin{eqnarray}
{\cal H} &=& \epsilon_0 \sum_{\langle i j\rangle} P_{i j}(0) +
\epsilon_1 \sum_{\langle i j\rangle} P_{i j}(1) + \epsilon_2
\sum_{\langle i j\rangle} P_{i j}(2), \label{Hamiltonian_N1b}
\\
\epsilon_0 &=& -4J_2+2J_1-J_0 ,\nonumber\\
\epsilon_1 &=& -J_2+J_1-J_0,\nonumber\\
\epsilon_2 &=& -J_2-J_1-J_0. \label{eviaJ}
\end{eqnarray}
Here $P_{i j}(S)$ is a projection operator for a pair of spins on
near neighbor sites $i$ and $j$ into a state with total spin
$S_i+S_j=S$ ($S=0,1,2$). Equivalence of (\ref{Hamiltonian_N1a})
and (\ref{Hamiltonian_N1b}) can be proven by noting simple
operator identities for two spin one particles
\begin{eqnarray}
1 &=& P_{i j}(0)+P_{i j}(1)+P_{i j}(2), \nonumber\\
(\vec{S}_i+\vec{S}_j)^2 &=& 4 + 2 \vec{S}_i \vec{S}_j
= 2 P_{i j}(1)+6 P_{i j}(2),\nonumber\\
(\vec{S}_i+\vec{S}_j)^4 &=& 16 + 16 \vec{S}_i \vec{S}_j + 4
(\vec{S}_i \vec{S}_j)^2 = 4 P_{i j}(1)+ 36 P_{i j}(2).
\end{eqnarray}
Note that states $|S_i=1,S_j=1;S_i+S_j=S\rangle$ have only the trivial
degeneracy corresponding to possible projections
of total spin $S$ on a fixed quantization axis $D_S=2S+1$.

Since we know a general form of our effective Hamiltonian, we can
compute $\epsilon_{0,1,2}$  by calculating the expectation values
of energy for arbitrary states in the appropriate subspaces
\begin{eqnarray}
\epsilon_S= - t^2\sum_m \frac{|<m|(a^{\dagger}_{i p}a_{j
p}+a^{\dagger}_{j p}a_{i p})|S_i=1,S_j=1;S_i+S_j=S>|^2}{E_m-E_0}.
\label{HSecondOrder}
\end{eqnarray}
Here $E_0=2U_2$ is the energy of the configuration with $N=2n+1$
bosons in each of the two wells, and $E_m$ is the energy of the
intermediate (virtual) states, $m$, that have  $2n$ and $2n+2$
bosons in the two wells, respectively. Both energies should be
computed in the zeroth order in $t$.

It is useful to note that the tunnelling Hamiltonian is spin
invariant; therefore, intermediate states  in $m$ summation in
(\ref{HSecondOrder}) should also have total spin $S$. Another
constraint on the possible states $m$ comes from the fact that the
tunnelling term can only change the spin on each site by $\pm 1$
since, in a Hilbert space of each well, operators $a^{\dagger}_{i
p}, a_{i p}$ act as vectors, according to their transformational
properties (\ref{avector1})-(\ref{avector2}).

 Direct calculations in Appendix \ref{AppendixA} give
\begin{eqnarray}
\epsilon_0 = -\frac{4t^2(n+1)(2n+3)}{3(U_0-2 U_2)}
-\frac{16t^2n(5+2n)}{15( U_0+ 4 U_2)}, \label{epsilon0}
\end{eqnarray}
\begin{eqnarray}
\epsilon_1 = -\frac{4t^2n(5+2n)}{5(U_0+4U_2)}, \label{epsilon1}
\end{eqnarray}
\begin{eqnarray}
\epsilon_2 = -\frac{28t^2n(5+2n)}{75(U_0+4U_2)}-\frac{4 (15 + 20 n
+ 8 n^2)}{15(U_0+U_2)}. \label{epsilon2}
\end{eqnarray}

Combining (\ref{Hamiltonian_N1b})-(\ref{epsilon2}) we find

\begin{eqnarray}
\frac{J_0}{t^2} &=& \frac{4 (15 + 20 n + 8 n^2)}{45 (U_0+U_2)}-\frac{4
(1+n)(3+2n)}{9 (U_0+2U_2)}+\frac{128(5+2 n)}{225 (U_0+4U_2)},
\nonumber\\
\frac{J_1}{t^2} &=& \frac{2(15 + 20 n + 8
n^2)}{15(U_0+U_2)}-\frac{16(5+2n)n}{75 (U_0+4U_2)},
\nonumber\\
\frac{J_2}{t^2} &=& \frac{2(15 + 20 n + 8 n^2)}{45
(U_0+U_2)}+\frac{4(1+n)(3+2 n)}{9(U_0-2U_2)}+
\frac{4n(5+2n)}{225(U_0+4U_2)}.
\label{Js}
\end{eqnarray}

\begin{figure}
\psfig{file=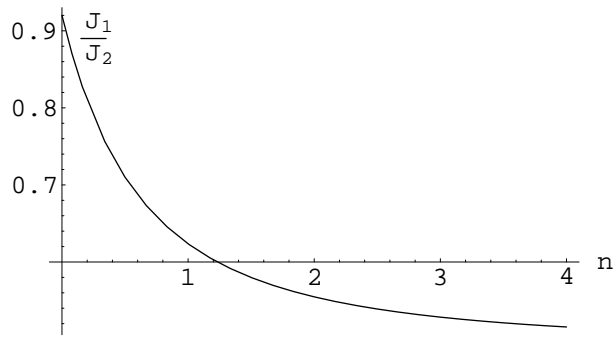} \caption{\label{j1j2} Ratio $J_1/J_2$ for
the effective spin Hamiltonian (\ref{Hamiltonian_N1a}) for an odd
number of bosons  $N=2n+1$ per site. $U_2=0.04 U_0.$}
\end{figure}

It will turn out that the ratio between $J_1$ and $J_2$ determines
magnetic ground state, and its dependence on $n$ is quite fast, as
shown in Fig. \ref{j1j2}.

 We now discuss limitations of the Hamiltonian
(\ref{Hamiltonian_N1a}) with (\ref{Js}). In the  insulating state
with exactly one boson per site, near neighbor interactions always
have the form (\ref{Hamiltonian_N1a}). Explicit expressions for
the $J$'s given in (\ref{Js}) only apply in the limit $t<<U_0$.
When $t$ becomes comparable to $U_0$ (but we are still in the
insulating phase), higher order terms become important, including
the possibility of spin coupling beyond the near neighbor sites.
In the insulating state with more than one boson per site
($N=2n+1$, $n>0$), we have an additional constraint: we should be
able to neglect configurations with spins on individual sites
higher than $1$. Matrix elements for scattering into such states
are of the order of $(Nt)^2/U_0$ (see (\ref{Js})), and their
energy is set by $U_2$. Therefore, the Hamiltonian
(\ref{Hamiltonian_N1a}) applies only when $Nt << (U_0 U_2)^{1/2}$,
which is well within the insulating state when $U_2<<U_0$ (SI
transition takes place for $Nt \sim U_0$).

\subsection{Phase diagram}
To understand the nature of the Hamiltonian
(\ref{Hamiltonian_N1a}) in the relevant regime of parameters
$J_2>J_1>0$, it is useful to start by considering a two-site
problem
\begin{eqnarray}
{\cal H}_{12} = -J_1 \vec{S}_1 \vec{S}_2 - J_2 ( \vec{S}_1
\vec{S}_2 )^2 \label{S=1TwoSites}
\end{eqnarray}
with $S_1=S_2=1$. Eigenstates of (\ref{S=1TwoSites}) can be
classified according to the value of the total spin $S_{tot}$, and
their energies may be computed using $2 \vec{S}_1 \vec{S}_2 =
S_{tot}(S_{tot}+1) -4$. Two spin one particles can combine into
$S_{tot}=0$, $1$, and $2$.
%\begin{eqnarray}
\begin{table}
\begin{tabular}{cccc}
{$S_{tot}$} & {$\vec{S}_1 \vec{S}_2$} & {$(\vec{S}_1
\vec{S}_2)^2$} &
{${\rm Energy}$} \\
\hline
0 & -2 & 4 & $2J_1-4J_2$\\
1 & -1 & 1 &$J_1-J_2$\\
2 & 1 & 1 & $- J_1 -J_2$
\end{tabular}
\caption{Eigenstates of a two site problem (\ref{S=1TwoSites})}
\label{TableS=1TwoSites}
\end{table}
The $J_1$ term in (\ref{S=1TwoSites}) favors maximizing $\vec{S}_1
\vec{S}_2$ by making the fully polarized $S_{tot}=2$ state. By
contrast, the $J_2$ term favors maximizing $(\vec{S}_1
\vec{S}_2)^2$ by forming a singlet state $S_{tot}=0$ (see Table
\ref{TableS=1TwoSites}). So, the latter term acts as an effective
antiferromagnetic interaction for this spin one system, and it
dominates for $J_2>J_1$. If we go beyond a two site problem and
consider a large lattice, we see that each pair of near neighbor
sites wants to establish a singlet configuration when $J_2 > J_1$.
However, because one cannot form singlets on two different bonds
that share the same site, some interesting spin order, whose
precise nature will depend on the lattice and dimensionality, will
appear.

\subsubsection{Phase diagram for $d=1$}

%%%
>From the discussion above we see the conflict intrinsic to the
Hamiltonian (\ref{Hamiltonian_N1a}): each bond wants to have a
singlet spin configuration, but singlet states on the neighboring
bonds are not allowed. There are two simple ways to resolve this
conflict:
\newline
A) Construct a state that mixes $S=0$ and $S=2$ on each bond but
can be repeated on neighboring bonds;
\newline
B) Break translational symmetry and favor singlets either on every
second bond.

At the mean-field level, solution of the type A is given by
\begin{eqnarray}
|N\rangle &=& \prod_i |S_i=1,m_i=0 \rangle .\label{NematicD1}
\end{eqnarray}
This can be established by noting that for any neighboring pair of
sites we indeed have a superposition of $S=0$ and $S=2$ states
\begin{eqnarray}
|S_i=1,m_i=0 \rangle |S_j=1,m_j=0 \rangle = -\frac{1}{\sqrt{3}}
|S_{tot}=0 \rangle + \sqrt{\frac{2}{3}} |S_{tot}=2, m_{tot}=0
\rangle. \label{NematicD1B}
\end{eqnarray}
State (\ref{NematicD1}) describes a nematic state that has no
expectation value of any component of the spin $\langle
S_i^{x,y,z}=0 \rangle$, but spin symmetry is broken since $\langle
(S_i^{x})^2 \rangle = \langle (S_i^{y})^2 \rangle = 1/2$ and
$\langle (S_i^{z})^2 \rangle = 0$. It is useful to point out the
similarity between wave function (\ref{NematicD1B})that mixes
singlet and quintet states on each bond, and a classical
antiferromagnetic state for spin $1/2$ particles that mixes spin
singlets and triplets on each bond.  Coleman's theorem
\cite{Coleman} (the quantum analog of Mermin-Wagner theorem)
forbids the breaking of spin symmetry in $d=1$, even at $T=0$.
However, a spin singlet gapless ground state that has a close
connection to the nematic state (\ref{NematicD1}) has been
proposed in \cite{Chubukov,S1EffectiveField} for $J_2$ close to
$J_1$.

The simplest way to construct a solution of type B is to take
\begin{eqnarray}
|D\rangle &=& \prod_{i=2n} |S_i=1,S_{i+1}=1,S_i+S_{i+1}=0 \rangle.
\label{DimerizedD1}
\end{eqnarray}
Such a dimerized solution has exact spin singlets for pairs of
sites $2n$ and $2n+1$, but pairs of sites $2n$ and $2n-1$ are in a
superposition of $S=0$, $1$, and $2$ states.

According to the variational wave functions (\ref{NematicD1}) and
(\ref{DimerizedD1}), the dimerized solution becomes favorable over
a nematic one only for $J_2/J_1 > 3/2 $ in $d=1$. However,
numerical simulations \cite{numeric1D} showed that for $J_2 >
J_1$, the ground state is always dimerized. It is a spin singlet
and has a gap to all spin excitations. This means that the
variational wave function (\ref{DimerizedD1}) may only be taken as
a caricature of the true ground state, although it captures such
key aspects of it, such as broken translational symmetry and the
absence of spin symmetry breaking.

\subsubsection{Phase diagram  for $d=2,3$}

The nematic state for the Hamiltonian (\ref{Hamiltonian_N1a}) in a
simple cubic lattice ($d=3$) for $J_2>J_1$ has been discussed
using mean-field calculations\cite{Chen1973},a semiclassical
approach\cite{Papanicolau}, and numerically \cite{numeric3D}.
Finally, recent work of Tanaka {\it et.al} \cite{Tanaka01}
provided a rigorous proof of the existence of the nematic order
at least in some part of this region, which satisfies  $2.66
J_1>J_2\ge 2 J_1$. The variational state for the nematic order may
again be given by equation (\ref{NematicD1}) and its mean field
energy is $E^{MF}_N=-2 J_2$. It is important to emphasize,
however, that the actual ground state is sufficiently different
from its mean-field version (\ref{NematicD1}). It is possible to
write down dimerized states with energy expectation lower than $-2
J_2$; however, numerical results \cite{numeric3D} suggest that the
ground state  doesn't break translational symmetry. The way to
obtain a more precise ground state wave function is to include
quantum fluctuations near the mean field state, as was done in
\cite{Papanicolau}. Hence, the mean-field wave function
(\ref{NematicD1}) does not provide a good approximation of the
ground state energy of the nematic state. Nevertheless, it is
useful for the discussion of order parameter and broken symmetries
of the nematic state.

In the nematic state, spin space rotational group $O(3)$ is
broken, though time reversal symmetry is preserved. The order
parameter for the nematic state is a tensor
\begin{eqnarray}
Q_{ab} = \langle S_a S_b \rangle -  \frac{ \delta_{ab} }{3}
\langle S^2 \rangle. \label{NematicQ}
\end{eqnarray}
In the absence of ferromagnetic order $\langle S_a S_b \rangle=
\langle S_b S_a \rangle$; hence, $Q_{ab}$ is a traceless symmetric
matrix. The minimum energy of (\ref{Hamiltonian_N1a}) is achieved
for $Q_{ab}$ that has two identical eigenvalues, which corresponds
to a uniaxial nematic \cite{deGennes}.

Then, the tensor $Q_{ab}$ can be written using a unit vector
$\vec{d}$ as
\begin{equation}
Q_{ab}=Q(d_a d_b-\frac13 \delta_{ab}). \label{Qvian}
\end{equation}
Vector $\vec{d}$ is defined up to the direction (i.e. $\pm
\vec{d}$ are equivalent) and corresponds to the director order
parameter \cite{deGennes}.  For the mean-field state
(\ref{NematicD1}), the director $\vec{d}$ can also be defined from
the condition that locally our system is an eigenstate of the
operator $\vec{d} \vec{S}$ with eigenvalue zero. However, such a
definition may not be applied generally.

The nematic phase behaves in many aspects as
antiferromagnetic\cite{AndreevGrischuk}, the direction of
$\vec{d}$ being analogous to staggered magnetization. Namely in
weak magnetic fields, $\vec{d}$ aligns itself in the plane
perpendicular to magnetic field, and spin-wave excitations have
linear dispersion\cite{S1EffectiveField}, with velocity
$$
c = \sqrt{2zJ_2(J_2-J_1)}.
$$
Nematic phases for the system of spin-1 particles have been
considered before in literature
\cite{Chubukov,Tanaka01,numeric3D,numeric1D,Papanicolau,AndreevGrischuk,Chen1973,S1EffectiveField},
so we will not discuss them here more extensively.

\section{ Insulating States with Two Atoms per Site}
\label{N2}

In this section we consider an insulating state of two bosons per
site. Possible spin values for individual sites are  $S=0$ and
$S=2.$ In the limit $t=0$, the interaction part of the Hamiltonian
(the $U_2$ term) is minimized when $S=0.$ The amplitude for
creating $S=2$ states, as well as the exchange energy of the
latter, is of the order of $t^2/U_0.$ So, when $t$ is of the order
of $(U_0 U_2)^{1/2}$ or larger, we may no longer assume that we
only have singlets in individual sites, and we need to include
$S=2$ configurations in our discussion. This regime is still
inside the insulating phase for small enough $U_2/U_0$ (the
superfluid-insulator transition takes place for $z t \sim U_0$).
In this section we will assume that $U_2/U_0$ is small enough, so
that $S=2$  becomes important in the insulating phase, before the
transition to superfluid. More careful consideration of superfluid
transition line and comparison with the case of $^{23} Na$ will be
presented in section \ref{GlobalPhase diagram}.
 In section
\ref{HeffS2} we exactly solve the problem for two wells. In
section \ref{latticehamilonian} we derive an effective Hamiltonian
that takes into account competition between spin gap of individual
sites, that favors $S=0$ everywhere, and exchange interactions
between neighboring sites that favor proliferation of $S=2$
states. Mean field solution of the effective magnetic Hamiltonian
is considered in section \ref{MeanFieldS2} and we find first order
quantum phase transition from isotropic to nematic phase. We
discuss collective excitations in sections
\ref{spinsingletexcitations} and \ref{nematicspinwaves} and the
effects of magnetic field in section \ref{magneticfieldeffects}.
We note that the state with $N=2$ has an advantage over states
with higher $N$ from an experimental point of view since it has no
three-body decays.

\subsection{Two site problem: exact solution}
\label{HeffS2}

To construct an effective magnetic Hamiltonian for this system, we
note that in the second order in $t$ it can be written as a sum of
interaction terms for all near neighbor sites (identical for all
pairs of sites). These pairwise interactions can be found by
solving a two well problem and finding the appropriate eigenvalues
and eigenvectors in the second order in $t$.

The Hilbert space for two sites with two atoms in each well is
given by the  direct sum of the following subspaces:
\begin{eqnarray}
|E_1>&=&|N_1=2,N_2=2,S_1=S_2=0,S_1+S_2=0>,
\nonumber\\
|E_2>&=&|N_1=2,N_2=2,S_1=S_2=2,S_1+S_2=0>,
\nonumber\\
|E_3>&=&|N_1=2,N_2=2,S_1=0,S_2=2,S_1+S_2=2>,
 \nonumber\\
|E_4>&=&|N_1=2,N_2=2,S_1=2,S_2=0,S_1+S_2=2>,
\nonumber\\
|E_5>&=&|N_1=2,N_2=2,S_1=S_2=2,S_1+S_2=2>,
\nonumber\\
|E_6>&=&|N_1=2,N_2=2,S_1=S_2=2,S_1+S_2=1>,
\nonumber\\
|E_7>&=&|N_1=2,N_2=2,S_1=S_2=2,S_1+S_2=3>,
\nonumber\\
|E_8>&=&|N_1=2,N_2=2,S_1=S_2=2,S_1+S_2=4>.
\label{TwoS2HilbertSpace}
\end{eqnarray}

Hopping term in (\ref{OriginalHamiltonian}) conserves total spin;
therefore, energy in each subspace doesn't depend on  the $z$
component of $S_1+S_2$, and states $|E_6>$, $|E_7>$, and $|E_8>$
form orthogonal subspaces that do not mix with any other states.
We can then use formula analogous to (\ref{HSecondOrder}) to
calculate corrections to the energies of these states in the
second order in $t$ (see Appendix \ref{AppendixB} for details):
\begin{eqnarray}
\tilde{\epsilon}_6 &=& 6 U_2 - 4\frac{t^2}{U_0},\nonumber\\
\tilde{\epsilon}_7 &=& 6 U_2 - 4\frac{t^2}{U_0},\nonumber\\
\tilde{\epsilon}_8 &=& 6 U_2 - 12\frac{t^2}{U_0}.
\label{epsilontilda68}
\end{eqnarray}
In  (\ref{epsilontilda68}) we used $U_2<<U_0$ and neglected $U_2$
relative to $U_0$ in denominators of the exchange terms.

\begin{figure} \psfig{file=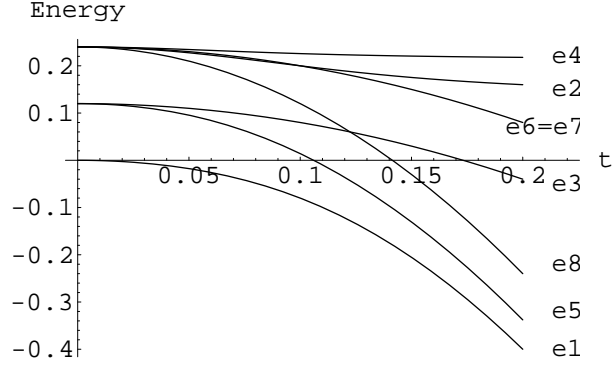} \caption{\label{e1e8}
Eigenstates of the effective spin Hamiltonian for a two site
problem with two atoms per site. Energy and $t$ are measured in
units of $U_0$, and we assumed $U_2=0.04 U_0 $. The lowest energy
states correspond to total spin $S=0$ ($e1$), $S=2$ ($e5$), and
$S=4$ ($e8$).}
\end{figure}

Boson tunnelling  can connect two subspaces, $|E_1>$ and $|E_2>$,
and three subspaces, $|E_3>$, $|E_4>$, and $|E_5>$ (only states
with the same component of $S^z$ have tunnelling matrix elements).
Thus, the energies and eigenstates should be found by
diagonalizing the matrix
\begin{eqnarray}
{\cal H}_{\beta\alpha} = E_{0\alpha} \delta_{\alpha\beta} -
t^2\sum_m \frac{<\beta|(a^{\dagger}_{1 p}a_{2 p}+a^{\dagger}_{2
p}a_{1p})|m> <m|(a^{\dagger}_{1 p}a_{2 p}+a^{\dagger}_{2
p}a_{1p})|\alpha>}{E_m-E_{0\alpha}}. \label{calH}
\end{eqnarray}
Here $E_{0\alpha}$ is the energy of the state $\alpha$ in the
zeroth order in $t$: $E_{01}=0$,
$E_{02}=6U_2$,$E_{03}=3U_2$,$E_{04}=3U_2$, $E_{05}=6U_2$.
Summation over $m$ covers intermediate states that have 1 and 3
bosons in the two wells, have the same total spin as states
$\alpha$ and $\beta$, and have spins in individual wells that
differ from $\alpha$ and $\beta$ by $\pm1$. The explicit
calculation presented in Appendix B gives the matrix ${\cal
H}_{\beta\alpha}$ and its eigenvalues. Energies of the states with
total spin zero are
\begin{eqnarray}
\tilde{\epsilon}_1 &=& 3U_2-\frac{6 t^2}{U_0}-
\sqrt{(3U_2-\frac{6t^2}{U_0})^2+
40 \frac{U_2 t^2 }{U_0}}, \nonumber\\
\tilde{\epsilon}_2 &=& 3U_2-\frac{6 t^2}{U_0}+
\sqrt{(3U_2-\frac{6t^2}{U_0})^2+ 40 \frac{U_2 t^2 }{U_0}}.
\label{Energystates12}
\end{eqnarray}
Energies of the states with $S=2$ are
\begin{eqnarray}
\tilde{\epsilon}_3 &=&3 U_2 - 4\frac{t^2}{U_0}  ,\nonumber\\
\tilde{\epsilon}_4 &=& \frac12\left(9 U_2-12t^2 +
\sqrt{144\frac{t^4}{U_0^2} + 40
\frac{t^2}{U_0} U_2 + 9U_2^2}\right)  ,\nonumber\\
\tilde{\epsilon}_5 &=& \frac12\left(9 U_2-12t^2 -
\sqrt{144\frac{t^4}{U_0^2} + 40 \frac{t^2}{U_0} U_2 +
9U_2^2}\right)  ,\label{Energystates345}
\end{eqnarray}
and each of these states is fivefold degenerate. Energies
$\tilde{\epsilon}_1$-$\tilde{\epsilon}_8$ are shown in FIG.
\ref{e1e8}.

For $U_2>0$ the lowest energy state is a total spin singlet that
has some mixture of $S=2$ states in individual wells when $t$ is
nonzero. The next favorable state has total spin $2$,
$|E_5\rangle$. When the value of $t$ is increasing, the
ferromagnetic state $|E_8\rangle$ becomes the third low lying
state.
 At this point, we have solved the problem for two wells, taking into account competition between
hopping and the $U_2$ term (overall Hilbert space for two wells is
$36$ dimensional).

\subsection{Effective Spin Hamiltonian for an Optical Lattice}
\label{latticehamilonian}

In the previous subsection we used perturbation theory in
tunnelling $t$ to study the problem of two sites with two atoms in
each well.  If we label the two sites $1$ and $2$, in the second
order in $t$ the effective Hamiltonian can be written as
\begin{eqnarray}
{\cal H}_{12} = 3 U_2 \left[ P(S_1=2)+P(S_2=2)\right]  +
 |\alpha>_1 |\beta>_2 {J}_{\alpha,\beta;\gamma
,\delta}<\gamma|_1<\delta|_2, \label{TwositeHamiltonianN2}
\end{eqnarray}
Here $P(S_{\{1,2\}}=2)$ are projection operators into states with
spin $S=2$ on sites $1$ and $2$ and ${J}_{\alpha,\beta;\gamma
,\delta}$ gives exchange interactions that arise from virtual
tunnelling processes into states with particle numbers
$(n_1=1,n_2=3)$ and $(n_1=3,n_2=1)$. The second term of
(\ref{TwositeHamiltonianN2}) includes all initial states ($|\gamma
\rangle_1$ and $|\delta \rangle_2$ for sites $1$ and $2$,
respectively) and all final states ($|\alpha \rangle_1$ and
$|\beta \rangle_2$).

Generalization of the effective spin Hamiltonian
(\ref{TwositeHamiltonianN2}) for the case of optical lattice is
obviously
\begin{eqnarray}
{\cal H} = 3 U_2 \sum_i P(S_i=2) + \sum_{<ij>} |\alpha>_i
|\beta>_j {J}_{\alpha,\beta;\gamma ,\delta}<\gamma|_i<\delta|_j,
\label{LatticeHamiltonianN2A}
\end{eqnarray}
This Hamiltonian is linear in $U_2$ and, therefore, can be written
as a sum of the bond terms
\begin{eqnarray}
{\cal H} &=& \sum_{\langle ij \rangle } \tilde{{\cal H}}_{ij},
\nonumber\\
\tilde{{\cal H}}_{ij} &=& \frac{ 3 U_2}{z} \left[
P(S_i=2)+P(S_j=2) \right] \nonumber\\&+& |\alpha>_i |\beta>_j
{J}_{\alpha,\beta;\gamma ,\delta}<\gamma|_i<\delta|_j.
\label{LatticeHamiltonianN2B}
\end{eqnarray}
Individual terms $\tilde{{\cal H}}_{ij}$ differ from
(\ref{TwositeHamiltonianN2}) only by rescaling $U_2 \rightarrow
\frac{U_2}{z}$, where $z$ is the coordination number of the
lattice. We did not give explicit expressions for
${J}_{\alpha,\beta;\gamma ,\delta}$ in the basis of eigenstates of
individual spins $S_i$ and $S_j$ but in the basis of eigenstates
of the total spin of the pair (\ref{TwoS2HilbertSpace}),
expressions for ${J}_{\alpha,\beta;\gamma ,\delta}$ can be
obtained from eigenstates and eigenvalues of
(\ref{TwositeHamiltonianN2}) (see equations (\ref{epsilontilda68})
-(\ref{Energystates345}) and Appendix B)(with a rescaled $U_2$).
Therefore, we can write
\begin{eqnarray}
\tilde{\cal H}_{ij} =
\sum_{\alpha,\beta,S^{\alpha}_z}|E^{ij}_{\alpha},S^{\alpha}_z>
\tilde{\cal  H}_{\alpha\beta}< E^{ij}_{\beta},S^{\alpha}_z|,
\label{LatticeHamiltonianN2C}
\end{eqnarray}
where states $|E^{ij}_{\alpha},S^{\alpha}_z>$
have been defined in equations (\ref{TwoS2HilbertSpace}).
Expressing $|E^{ij}_{\alpha},S_z>$ via states
$|N_i=2,S_i=\{0,2\},S_{i z}=-S_i...S_i>$ using known
Clebsch-Gordon coefficients
$$
|E^{ij}_{\alpha},S_z>= \sum_{S^{\alpha}_{i z} S^{\alpha}_{j z}}
C^{S^{\alpha},S^{\alpha}_z}_{S^{\alpha}_i, S^{\alpha}_{i z
};S^{\alpha}_j, S^{\alpha}_{j z
}}|N_i=2,S^{\alpha}_i,S^{\alpha}_{i
z}>|N_j=2,S^{\alpha}_j,S^{\alpha}_{j z}>,
$$
we can write the Hamiltonian (\ref{LatticeHamiltonianN2A})
as
\begin{equation}
 \hat H = \sum_{<ij>} |\alpha>_i |\gamma>_j
{\cal H}_{\alpha,\beta;\gamma ,\delta}<\beta|_i<\delta|_j,
\label{abgd}
\end{equation}
where states $|\alpha \rangle$ - $|\delta\rangle $ belong to the
set $\{S=0\},\{S=2,S_z=-2,...,2\}$, and ${\cal
H}_{\alpha,\beta;\gamma ,\delta}$  is given by proper rotation of
$\tilde{\cal H}_{\alpha,\beta}.$

\subsection{Phase Diagram from the Mean-Field Calculation}
\label{MeanFieldS2}

In this section we study the phase diagram of the system described
by the Hamiltonian (\ref{abgd}) using translational invariant
variational wave functions. Such mean-field approach gives correct
ferromagnetic and antiferromagnetic states for Heisenberg
hamiltonians in $d \geq2$, so we expect it to be applicable in our
case. We think that this approach successfully captures the main
features of the system: first order transition between the spin
gapped and the nematic phases, the nature of the order parameter
in the nematic phase, and elementary excitations in both phases.
However, we cannot rule out the possibility of more exotic phases
that fall outside of our variational wave functions, for example,
the dimerized phase discussed in \cite{zhou02}, and numerical
calculations are required to study if such phases will actually be
present.

 As we saw in the previous chapter, energy of the
two-well problem is minimized when total spin is $0.$ However,
energy on all bonds cannot be minimized simultaneously, so we
cannot solve a problem exactly for a lattice. We use a mean field
approach to overcome this difficulty, taking variational wave
function
\begin{eqnarray}
|\Psi>=\prod_i(c_{0,0}|N_i=2,S=0,S_z=0>+\sum_{
m=-2,...,2}c_{2,m}|N_i=2,S=2,S_z=m>),
\label{S2variationalWavefunction}
\end{eqnarray}
\begin{equation}
|c_{0,0}|^2+\sum_{m=-2,...,2}|c_{2,m}|^2 =1.
\label{cnormalization}
\end{equation}
Now we can evaluate expectation value of energy over variational
state (\ref{S2variationalWavefunction}) and find the ground state
numerically.
%Details of this evaluation and
%numerical procedures described in Appendix \ref{AppendixC}, here
%we only state the results.

We parameterize~(\ref{cnormalization}) as
\begin{eqnarray}
c_{0,0} &=& \cos \theta, c_{2,m} = \sin\theta a_m,
\label{S2parametrization}
\end{eqnarray}
$$
\sum_{m=-2,...,2}|a_m|^2 =1.
$$
In Appendix \ref{AppendixC} we demonstrate that for a region of
$\theta$ where the mean-field energy  is minimized,  $[ a_m ]$ has
the form, up to $SU(2)$ rotations \footnote{Expression (\ref{am})
describes an eigenstate of $S^z$ with eigenvalue zero. After an
$SU(2)$ rotation we will have $[a_m]$ that is an eigenstate of
$\vec{d}\vec{S}$ with zero eigenvalue. Vector $\vec{d}$
corresponds to the direction of uniaxial nematic.},
\begin{eqnarray}
[ a_m ] =  (0,0,1,0,0)^{T}. \label{am}
\end{eqnarray}

Mean field energy does not depend on $\vec{d}$ and we find in the
region of interest the energy per lattice site to be
\begin{eqnarray}
E[\theta] =3 U_2\sin^2{\theta} + \frac{z t^2}{12 U_0}( - 51 +
4\cos{2\theta} +
        7 \cos{4\theta}-8\sqrt{2} \sin{2\theta} +
        4\sqrt{2}\sin{4\theta}).
\label{Etheta}
\end{eqnarray}
One can immediately see that if we try to expand this expression
near $\theta=0,$ there is no linear term, but second and third
order terms are present, which indicates that by changing the
parameters of the Hamiltonian, we will have a  first order quantum
phase transition, at which the value  of $\theta$ that minimizes
the energy changes discontinuously. This is typical for ordinary
nematics\cite{deGennes} since in Landau expansion third order
terms  are not forbidden by $\vec d \rightarrow - \vec d$
symmetry. The reason why our transition is of the first order can
be traced back to the fact that mean field energy has terms which
mix $c_{0,0}$ and $c_{2,m}$ in odd powers, i.e.
$c_{0,0}c_{2,-2}(c^*_{2,-1})^2,$ and overall $U(1)$ symmetry
doesn't prohibit odd powers of $\theta$ in (\ref{Etheta}).

\begin{figure}
\psfig{file=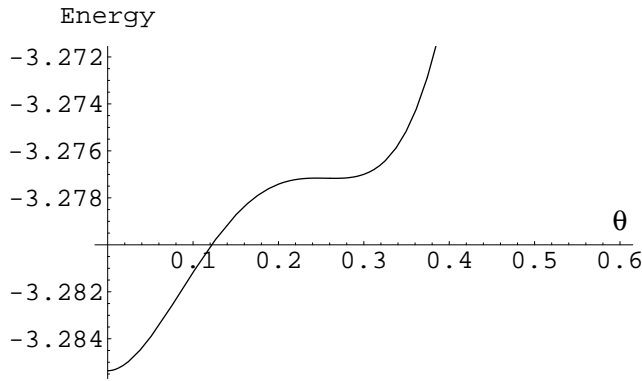} \caption{\label{energy1}Dependence of
the energy functional (\ref{Etheta}) on $\theta$
when $z
t^2/(U_0 U_2)\approx 0.4928$
(energy is given per lattice site in units of $U_2$).}
\end{figure}

\begin{figure}
\psfig{file=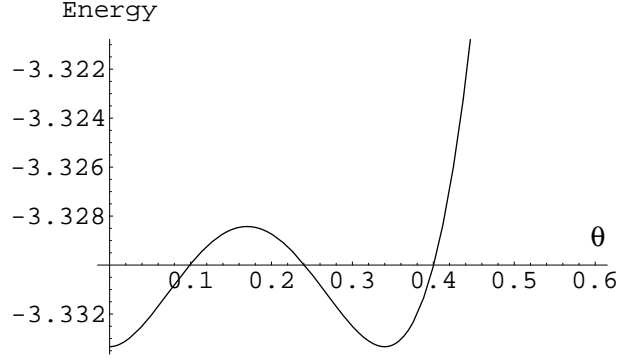} \caption{\label{energy2} Dependence of
the energy functional (\ref{Etheta}) on $\theta$
when $z
t^2/(U_0 U_2) =1/2$ (energy is given per lattice site per lattice site in units of $U_2$).}
\end{figure}

\begin{figure}
\psfig{file=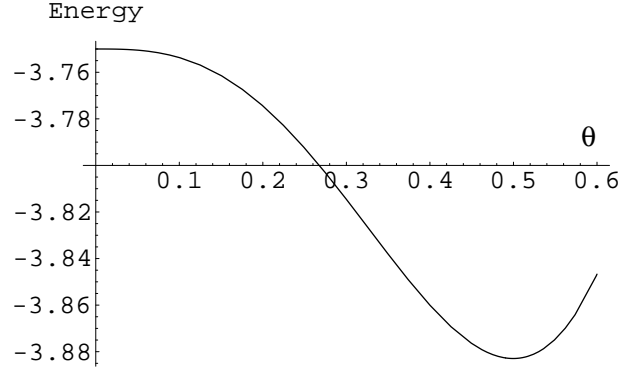} \caption{\label{energy3} Dependence of
the energy functional (\ref{Etheta}) on $\theta$ when $z t^2/(U_0
U_2)=9/16=0.5625$ (energy is given  per lattice site in units of
$U_2$).}
\end{figure}

Since phase transition is of the first order, transition is
characterized  by several regimes. First, when $t$ is small,
global  energy minimum is at $t=0$ and there are no other local
minima, i.e. we have spin singlets in all individual sites. Then,
when condition $z t_-^2/(U_0 U_2)\approx 0.4928$ is satisfied, a
local minimum appears at $\theta_-\approx0.25,$ see Fig.
\ref{energy1}. As we continue increasing $t$, the minimum at
nonzero $\theta$ becomes deeper, and eventually at $ z t_c^2/(U_0
U_2) = 1/2$ the global minimum of (\ref{Etheta}) is reached for
$\sin \theta_c = 1/3,$ (see Fig.\ref{energy2}).  However, there is
still a local minimum at $\theta=0.$ If we keep increasing $t$,
the minimum at $\theta=0$ becomes completely unstable at $z
t_+^2/(U_0U_2) = 9/16$ and there is only one minimum at
$\theta_+\approx0.5$ (see Fig. \ref{energy3}). As we increase $t$
further, $\sin \theta_+$ continues to grow, approaching the value
$\sin \theta_\infty = (2/3)^{1/2}$. It is useful to point out that
when $t$ is changed in experiments (e.g. by changing the strength
of the optical potential \cite{greiner02}), we expect that the
system will not switch between the singlet and nematic phases at
$t_c$, but will remain in the appropriate metastable local minimum
until it becomes completely unstable. So, in experiments with
increasing $t$, the transition from the singlet to the nematic
states will occur at $t_+$, and in experiments with decreasing
$t$, the transition from the nematic to the singlet state will
take place at $t_-$.  We note, however, that the difference
between different $t$ is quite small.

\begin{figure}
\psfig{file=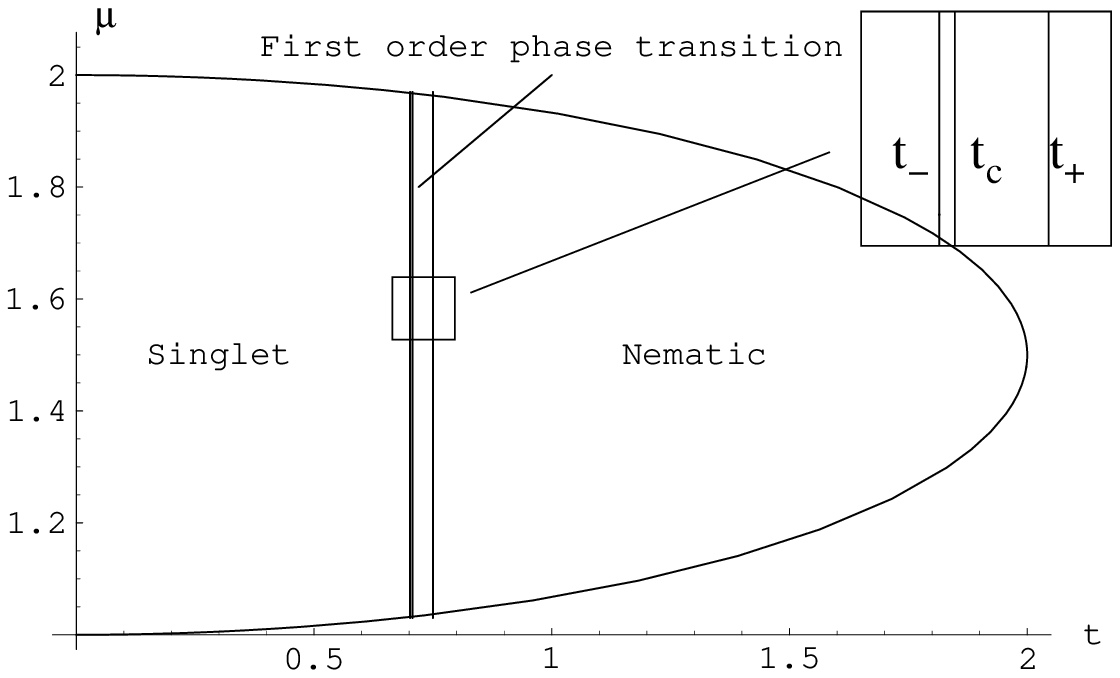} \caption{\label{N2diagram} Mott phase
diagram for $N=2$. $\mu$ is measured in units of $U_0$, $t$ is
measured in units of $\sqrt{\frac{U_0 U_2}{z}}$. $t_c$ marks the
actual first order phase transition and $t_-$ and $t_+$ correspond
to limits of metastability. Superfluid-Insulator transition line
is presented as an eye guide, and will be discussed in section
\ref{GlobalPhase diagram}.}
\end{figure}

In Figure \ref{N2diagram} we show the phase diagram for the
insulating phase with $N=2$, including a true first order
transition line at $t_c$ and limits of metastability at $t_-$ and
$t_+$. The Superfluid-Insulator transition line is shown as an eye
guide for the case of small enough $U_2/U_0$; its exact position
will be presented in section \ref{GlobalPhase diagram}.
 It is useful to point out that in the
discussion above we used canonical ensemble (fixed number of
particles) rather than grand canonical ensemble (fixed chemical
potential) to discuss the singlet to nematic phase transition.
However, intermediate states that contribute to exchange
interactions always involve one particle and one hole. Hence,
their energy does not depend on the chemical potential. This
explains why the singlet to nematic phase boundary in Figure
\ref{N2diagram} does not depend on $\mu$. It is consistent with
our physical intuition that insulating states have a certain
number of particles, but their chemical potential $\mu$ is not
well defined as long as $\mu$ is inside the Mott gap. In the
discussion presented in this section, we assumed that the system
remains  deep in the insulating phase and the superfluid to
insulator transition does not preempt the isotropic to nematic
transition inside the insulating lobe. Precise conditions under
which this is justified will be given in Section \ref{GlobalPhase
diagram}.

\begin{figure}
\psfig{file=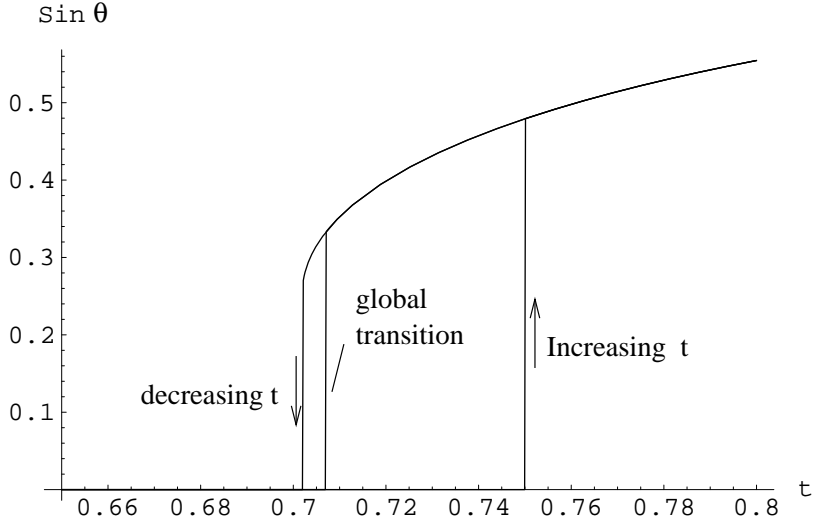} \caption{\label{thetaont}Dependence of
$\sin{\theta}$ in equation (\ref{Etheta}) on $t$, measured in
units of $\sqrt{\frac{U_0 U_2}{z}}$.}
\end{figure}

\subsection{Quantum Fluctuations Corrections for
the Spin Singlet State} \label{spinsingletexcitations}

For small enough $t$, mean-field analysis of the previous section
predicts the singlet  ground state that does not depend on $t$.
Now we will consider quantum fluctuations near this state to
obtain more accurate wave function and excitation spectra. We can
rewrite (\ref{abgd}) via Hubbard operators
\begin{eqnarray}
A^{\alpha \beta}_i = |\alpha\rangle_i \langle \beta |_i.
\label{Hubbard}
\end{eqnarray}
Here $|\alpha\rangle_i$ and $|\beta\rangle_i $
belong to the set $\{S=0\},\{S=2,S_z=-2,...,2\}$.
Commutation relations between $A_i^{\alpha \beta}$ are very simple:
\begin{equation}
[A_i^{\alpha \beta},A_i^{\gamma \delta}]=\delta_{\beta \gamma}
A_i^{\alpha \delta}-\delta_{\alpha \delta}A_i^{\gamma \beta}.
\label{Acomm}
\end{equation} Now we introduce boson operators $b^{\dagger}_{\alpha i}$
 that create states  with $S_i=2,S_{i z}=\alpha,$ and $c^{\dagger},$
that creates a singlet on $i$th site. Our physical subspace is
smaller than generic Fock space of these bosons and should satisfy
the condition
\begin{equation}
  c_i^{\dagger}c_i + \sum_{\alpha}b^{\dagger}_{i\alpha}b_{i\alpha}=1
  \label{constr}
\end{equation}
on each site.

 One can easily check that if we set $A^{\alpha
\beta}=b^{\dagger}_{\alpha} b_{\beta} $ for spin $S=2$ states and
similar substitution with $c$ bosons when one of the states is a
singlet state(which we will denote as $s$), then commutation
relations (\ref{Acomm}) are satisfied. Since for small enough $t$
only a singlet state is occupied in mean field approximation, we
can resolve  constraint (\ref{constr}) using analog of
Holstein-Primakoff representation near $c^{\dagger}c=1$ state
\cite{Papanicolau}, which is given by
\begin{eqnarray}
A_i^{\alpha \beta}=b^{\dagger}_{i\alpha} b_{i\beta},
A_i^{s \alpha}=(1-b^{\dagger}_{i\beta}b_{i\beta})^{1/2}b_{i\alpha}, \\
A_i^{s s}=1 - b^{\dagger}_{i\beta} b_{i\beta}, A_i^{\alpha
s}=b^{\dagger}_{i\alpha}(1-b^{\dagger}_{i\beta}b_{i\beta})^{1/2}.
\label{bosonization}
\end{eqnarray}
Now we expand our initial Hamiltonian in terms of now independent
operators $b^{\dagger}_i$ up to the second order: \begin{eqnarray}
\hat H^{(2)} = \sum_{<ij>} (H_{\alpha\beta;ss}b^{\dagger}_{\alpha
i }b^{\dagger}_{\beta j}+ H_{ss;\alpha\beta}b_{\alpha i}b_{\beta
j}+H_{\alpha s;s \beta}b^{\dagger}_{\alpha i}b_{\beta j} + H_{s
\alpha;
\beta s}b^{\dagger}_{\alpha j}b_{\beta i})+ \nonumber \\
+\frac{z}{2} \sum_{i}( H_{\alpha s; \beta s}b^{\dagger}_{\alpha
i}b_{\beta i}+H_{s \alpha;s \beta}b^{\dagger}_{\alpha i}b_{\beta
i}+H_{s s;s s}(1-b^{\dagger}_{\alpha i}b_{\alpha
i}-b^{\dagger}_{\alpha j}b_{\alpha j})). \label{H2b}
\end{eqnarray}
Calculation of matrices $H_{\alpha\beta;\gamma\delta}$ gives
necessary matrix elements
$$
H_{\alpha s;\beta s}=H_{s \alpha ;s \beta}= \delta_{\alpha \beta
}(-\frac{20}{3} \frac{t^2}{U_0}  + \frac{3 U_2}z),
$$
$$
 H_{s s;s s}=-\frac{20}3 \frac{t^2}{U_0},
$$
$$
 H_{\alpha s;s \beta}=H_{s \alpha ;\beta s}=-\delta_{\alpha \beta }\frac83 \frac{t^2}{U_0},
$$
$$
H_{\alpha \beta;s s}=H_{s s;\alpha \beta}=-\frac83 \frac{t^2}{U_0}
\left[\begin{array}{ccccc}
  0 & 0 & 0 &  0 & 1\\
  0 & 0 & 0 &  -1 & 0 \\
  0 & 0 & 1 &  0 & 0 \\
  0 & -1 & 0 &  0 & 0\\
  1 & 0 & 0 &  0 & 0
\end{array}\right].
$$
We rewrite our Hamiltonian in terms of Fourier transforms of
$b^{\dagger}_{\alpha k} ,b_{\alpha k}$:
\begin{eqnarray} \hat H^{(2)}
=E_0 + \frac{z}2 \sum_{k}
\gamma_k(H_{\alpha\beta;ss}b^{\dagger}_{\alpha k
}b^{\dagger}_{\beta - k}+ H_{ss;\alpha\beta}b_{\alpha k}b_{\beta -
k} + H_{\alpha s;s \beta}b^{\dagger}_{\alpha k}b_{\beta k} + H_{s
\alpha;
\beta s}b^{\dagger}_{\alpha k}b_{\beta k})+ \nonumber \\
+2(H_{s \alpha;s \alpha}-H_{s s;s s})b^{\dagger}_{\alpha
k}b_{\alpha k}, \label{H2k}
\end{eqnarray}
where  $E_0$ is classical energy and
\begin{equation}
\gamma_k=\frac1z\sum_{e_\mu}e^{i{\bf k e_\mu}}. \label{gammak}
\end{equation}
 Now we  will use canonical Bogoliubov transformations
to diagonalize this Hamiltonian. Since most of the terms are
diagonal in $\alpha,\beta$ subspace, it is easy to see that
required transformation mixes  operators $\{b^{\dagger}_{0 k},b_{0
-k}\},$ $\{b^{\dagger}_{1 k},b_{-1 -k}\},$ $\{b^{\dagger}_{-1
k},b_{1 -k}\},$ $\{b^{\dagger}_{2 k},b_{-2 -k}\},$ and
$\{b^{\dagger}_{- 2 k},b_{2 -k}\}.$

Transformation that mixes the first pair is
$$
b_{0 k}=\cosh{\theta_k} \beta_{0 k} + \sinh {\theta_k}
\beta^{\dagger}_{0 -k},
$$
$$
b_{0 -k}=\cosh{\theta_k} \beta_{0 -k} + \sinh {\theta_k}
\beta^{\dagger}_{0 k},
$$
and complex conjugates. Substituting this  transformation into
(\ref{H2k}) and requiring that terms with $\beta^{\dagger}_{0
k}\beta^{\dagger}_{0-k}$ and $\beta_{0 k}\beta_{0-k}$ vanish, we
obtain the equation for $\theta_k$
$$
\tanh{2 \theta_k} = -\frac{g_k}{f_k}=-\frac{\frac83\frac{z
t^2}{U_0}\gamma_k}{
 3 U_2-\frac{8}{3} \frac{z t^2}{U_0}\gamma_k},
$$
where
$$
g_k=\frac83\frac{z t^2}{U_0}\gamma_k,
$$
$$
f_k=3 U_2-\frac{8}{3}\frac{z t^2}{U_0}\gamma_k.
$$
Energy of this excitation becomes
\begin{eqnarray}
E(k)=\sqrt{f_k^2 - g_k^2 }=\sqrt{U_2(9 U_2-16\frac{z
t^2}{U_0}\gamma_k)}.
\label{modeEnergy}
\end{eqnarray}
\begin{figure}
\psfig{file=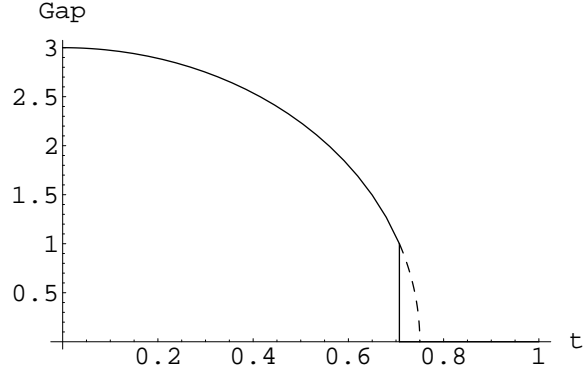} \caption{\label{gap}Dependance of excitation
gap(measured in units of $U_2$) on $t$, measured in units of
$\sqrt{\frac{U_0 U_2}{z}}.$}
\end{figure}

Equation (\ref{modeEnergy}) suggests that the  first instability
appears at $k=0$ and gives the phase boundary that agrees with the
metastability line $t_+$ found in the previous subsection.
However, results of the previous subsection suggest that phase
transition is of  the first order and takes place before the mode
softening at $k=0$. The first order transition may also be
obtained with the formalism presented in this section by noting
that expansion of (\ref{bosonization}) allows third order terms
$cb^{\dagger} b^{\dagger}+{\rm c.c.}$.

We can use similar analysis to discuss excitations with other spin
quantum numbers. For example, excitations
 with $S_z=\{+1,-1\}$ are
diagonalized by analogous Bogoliubov transformations with
$\theta_k \rightarrow -\theta_k,$ and excitations with
$S_z=\{+2,-2\}$ are diagonalized with transformations with the
same $\theta_k.$ As required by the spin symmetry of the singlet
state, all of these excitations have the same energy.

Now we can discuss the approximations made while expanding over
$b^{\dagger}_{\alpha k}, b_{\beta k}.$ While transformation
(\ref{bosonization}) is the exact resolution of the constraint
(\ref{constr}), expansion to the second order adds states with
higher boson occupation numbers and changes Hilbert space (this is
completely analogous to usual antiferromagnet spin-wave theory).
However, if {\it a posteriori}  we can verify that only states
with occupation numbers $n_{\alpha i}=\{0,1\}$ are present in the
ground state, then expansion of the constraint (\ref{constr}) up
to the second order was justified.
 The parameter that controls such expansion is
$$
 b^{\dagger}_{\alpha i} b_{\alpha i}=\frac1N\sum{ b^{\dagger}_{\alpha k} b_{\alpha
 k}}=\int{\sinh^2{\theta_k}\frac{d^dk}{(2\pi)^d}}.
$$
Calculation of this quantity while the singlet state is  still a
global maximum for $d=3$ gives numerical values $<0.001;$
therefore, our expansion is much more precise than for Heisenberg
antiferromagnet, where this quantity is not much smaller than $1$
and one needs the condition $S \gg 1$ to justify the spin wave
theory.

\subsection{Spin Wave Excitations in the Nematic Phase}
\label{nematicspinwaves}

\begin{figure}
\psfig{file=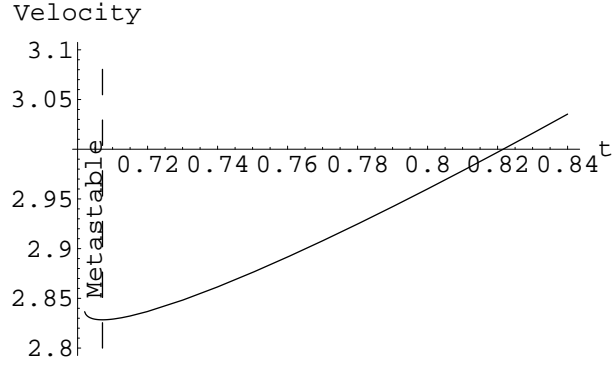} \caption{\label{velocity}Dependance of
the spin wave velocity (measured in units of $U_2/\sqrt{z}$) on
$t$, measured in units of $\sqrt{\frac{U_0 U_2}{z}}.$}
\end{figure}

Now we will consider excitations for the states with nematic
order. For the states described by
(\ref{S2variationalWavefunction}) - (\ref{am}), there are no
expectation values of the spin operators $\langle \vec{S} \rangle
=0$, but there is a nematic order (\ref{NematicQ}) when $\theta
\neq 0$. For example, when $\vec{d}$ is pointing along $z$ we find
\begin{eqnarray}
Q_{ab} = \sin^2 \theta \left( \begin{array}{ccc}
1 & 0 & 0 \\
0 & 1 & 0 \\
0 & 0 & -2
\end{array} \right)
\label{Qab}
\end{eqnarray}
In the singlet phase, expectation values of both the spin
operators are $\langle \vec{S} \rangle =0$ and $\langle Q_{ab}
\rangle =0$. In the singlet phase the system has a gap to all
excitations of order $U_2$, while  nematic phases have gapless
spin-wave excitations that originate from the continuous symmetry
breaking. The general form of the state with minimum energy is
expressed via Euler angles  of order parameter $\vec d$ as
$$
U_z(\gamma)U_y(\beta)U_z(\alpha) \{0,0,1,0,0\}^T,
$$
where $U(\gamma)$ are finite angle  rotation matrices. From
(\ref{Qab}) and (\ref{Qvian})  we can express the nematic order
parameter for such a state as
$$
Q_{a b}=-3\sin^2{\theta}(d_a d_b - \frac13\delta_{a b}).
$$
Goldstone theorem tells us that low lying modes will be
fluctuations of direction of $\vec d$, and there will be two
degenerate modes. We can utilize the approach used in the previous
subsection to consider excitations in the nematic phase. Here, we
should make generalized  Holstein-Primakoff expansion near the
nematic state. First, we make unitary transformation in Hilbert
subspace of each site, which is given by
\begin{equation}
 \left(
\begin{array}{c}
  |0\rangle \\
  |1\rangle \\
|2\rangle \\
  |3\rangle \\
  |4\rangle \\
  |5\rangle
\end{array}\right) =
\left(\begin{array}{cccccc}
  \cos \theta & 0 & 0 & \sin \theta & 0 & 0 \\
  0 & 0 & 1/\sqrt2 & 0 & 1/\sqrt2 & 0 \\
  0 & 0 & -1/\sqrt2 & 0 & 1/\sqrt2 & 0 \\
 - \sin \theta & 0 & 0 & \cos \theta & 0 & 0 \\
  0 & 1/\sqrt2 & 0 & 0 & 0 & 1/\sqrt2 \\
  0 & - 1/\sqrt2 & 0 & 0 & 0 & 1/\sqrt2
\end{array}\right)
\left(\begin{array}{c}
  |S=0,S=0\rangle \\
  |S=2,S_z=-2\rangle \\
|S=2,S_z=-1\rangle \\
  |S=2,S_z=0\rangle \\
  |S=2,S_z=1\rangle \\
  |S=2,S_z=2\rangle
\end{array}\right).
\label{Smatrix}
\end{equation}
Making appropriate transformation on
$H_{\alpha\beta;\gamma\delta},$ we can write our Hamiltonian as
\begin{equation}
 \hat H = \sum_{<ij>} |\alpha>_i |\gamma>_j
{\tilde {\cal  H}}_{\alpha,\beta;\gamma
,\delta}<\beta|_i<\delta|_j, \label{abgd2}
\end{equation}
where states $|\alpha \rangle$ - $|\delta\rangle $ belong to the
set $\{|0\rangle-|5\rangle\}.$ After that, we proceed exactly as
in the previous subsection, expanding near $|0\rangle$ state.
 Since dependance $\theta$ on $t^2/(U_0 U_2)$ is determined by the minimization
of the  energy, linear terms in $b_{k\alpha}$ and
$b_{k\alpha}^{\dagger}$ are absent. Quadratic terms have exactly
the same form as in (\ref{H2b}), and all matrices become diagonal
due to the proper basis choice (\ref{Smatrix}). Now we can use
Bogoliubov transformation to diagonalize the quadratic part. For
excitations to states $|1\rangle$ and $|2\rangle$, we obtain
energy dependance
$$
 E^2_1(k)=E^2_2(k)=\frac{1}{36} (-16 \gamma_k^2
\frac{z^2 t^4}{U_0^2} (4\cos{2 \theta} + \sqrt2\sin{2\theta})^2 +
$$
$$
(9\frac{z t^2}{U_0}
 - 18 \gamma_k \frac{z t^2}{U_0} + 9U_2 + (2(\gamma_k - 1 ) \frac{z t^2}{U_0} + 9U_2)\cos{2 \theta} - 7
\frac{z t^2}{U_0} \cos{4 \theta}
$$
$$
+ 4\sqrt2(1-\gamma_k)\frac{z t^2}{U_0} \sin{2 \theta}  - 4
\sqrt2\frac{z t^2}{U_0} \sin{4\theta})^2),
$$
where $\gamma_k$ was defined in (\ref{gammak}), and dependance of
$\theta$ on $z t^2/(U_0 U_2)$ is shown in Fig. \ref{thetaont}. We
find that for $k=0$, energies of these excitations are zero, as
expected for nematic waves from Goldstone theorem. These
excitations create states with $S_z=\pm 1.$ For small $\vec k$,
the energy of excitations depends linearly on $|\vec k|,$ and
dependance of spin wave velocity on the parameters of the lattice
is shown in Fig. \ref{velocity}.

\begin{figure}
\psfig{file=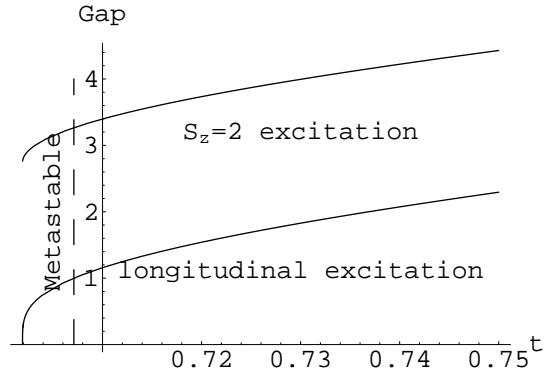} \caption{\label{gap2} Energy of the gapped
excitations in the nematic phase of $N=2$ atoms per site at zero
wavevector. The energy gaps and $t$ are measured in units of $U_2$
and $\sqrt{\frac{U_0 U_2}{z}},$ respectively.}
\end{figure}

Let us now consider gapped excitations for the nematic phase.
Excitation to the state $|3\rangle$ corresponds to longitudinal
fluctuations in the value of $\theta,$ and the energy of such
excitations becomes zero at $t_-$ since at this point fluctuations
of $\theta$ are not suppressed. Excitations to the states
$|4\rangle$ and $|5\rangle$ correspond to the creation of $S_z=\pm
2$ states and they are degenerate. For all of these excitations,
energies are minimized for $\vec k=0.$ Dependence of the gap on
parameters is shown in Fig. \ref{gap2}.

\subsection{Effects of Small Magnetic Field}
\label{magneticfieldeffects}

\begin{figure}[htbp]
\psfig{file=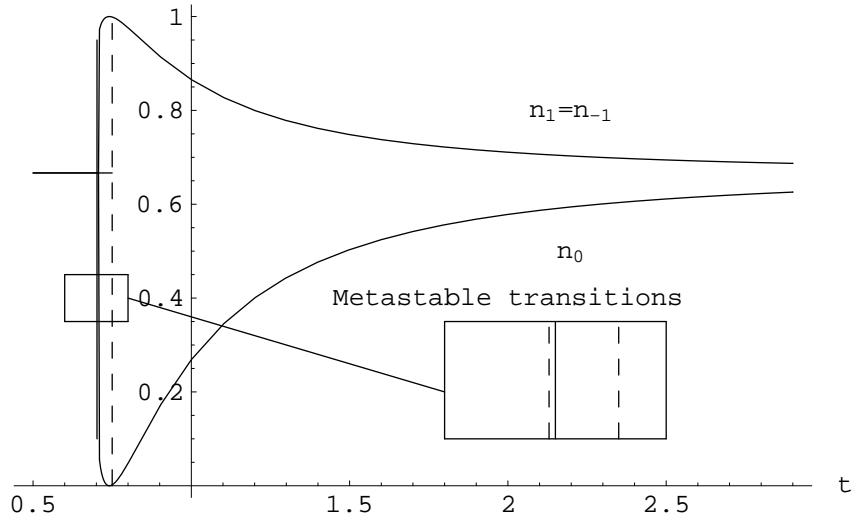} \caption{\label{n0n1}Dependence of
occupation numbers $n_0$ and  $n_{1}=n_{-1}$ on $t$ for an
insulating state with two bosons per site, $N=2$. Tunnelling $t$
is measured in units of $\sqrt{\frac{U_0 U_2}{z}}$.}
\end{figure}

Let us now consider the effect of a small magnetic field, $\mu H
<< U_2 $, on our system. For $U_2$ in the range of kHz (see Fig.
\ref{tU2}), this corresponds to magnetic fields smaller than a $1
mG $. We suppose that the field is small enough that it does not
change the scattering lengths due to the energy level shifts
inside of atoms. Since all atoms have the same gyromagnetic ratio,
interaction with external magnetic field depends only on the total
spin, and the internal structure of the states is not important.
%In the first order in $H$ the
%magnetic field has  no effect on the singlet state.
In the case of a nematically ordered insulating state, the ground
state energy doesn't have any contributions linear $H$ (this
follows from $<0|{\bf H S }|0>=0$), and the second order
contribution depends on the relative orientation of the nematic
order parameter $\vec{d}$ and magnetic field ${\vec H}$. Suppose
$\vec{d}$ is directing along the $z$ axis and $\vec{H}$ lies in
$x,z$ plane. In the second order perturbation theory, energy
correction to the ground state is always non positive:
$$
E^{(2)}=-\sum_{E_n}\frac{<0|\mu{\bf H S} |n><n|{\bf \mu H S
}|0>}{E_n-E_0}=-\sum_{E_n}(\mu
H_x)^2\frac{<0|S_x|n><n|S_x|0>}{E_n-E_0}<0; ,
$$
this quantity is of order $ -(\mu H_x)^2/t^2.$
 Since for the state with $\vec d||\vec H$ it is again zero and the system
doesn't benefit from magnetic field, energy is minimized when
$\vec d$ lies in a plane perpendicular to $\vec H$(this is
completely analogous to Antiferromagnet). Using this property, one
can a distinguish nematic phase  from a singlet phase. One should
apply a small magnetic field in $z$ direction to fix the plane in
which $\vec{d}$ lies, release the trap, and let the atoms fall in
the gravitational field with some magnetic gradient, to separate
the states with different $S_z$. Then, one should measure
quantities of each spin component. These values will have a sharp
change when we cross the first order phase transition line.
Knowing how to express spin states via original boson operators,
we can calculate expectation values  of different spin components
to be:
$$
n_{1}=n_{-1}=\frac23 \cos[\theta]^2 +
    \frac{\sqrt{2}}3 \cos[\theta]\sin[\theta] + \frac56
    \sin[\theta]^2,
$$
$$
n_{0}=\frac23 \cos[\theta]^2 -
    \frac{2\sqrt{2}}3 \cos[\theta]\sin[\theta] + \frac13
    \sin[\theta]^2.
$$
Using known expressions for dependence of $\theta$ on $t$, we can
make mean-field predictions on occupation numbers, shown in FIG.
\ref{n0n1}

\section{ Large number of particles per site}
\label{Nbig}

In this section we discuss the case  $ N\gg 1$ for both parities
of $N$. We show how one can separate variables describing angular
momentum  and the number of particles in each
well\cite{itpreport}, and derive an effective Hamiltonian which is
valid under conditions $U_2 , N t \ll U_0,$ which is less
restrictive than in section \ref{Nodd} for $N>>1.$

When we have $N$ spin-one bosons localized in a well in the same
orbital state, their total spin may take any value that satisfies
constraints
\begin{eqnarray}
S+N={\rm even},
\label{SNeven} \\
S \leq N. \label{SlessN}
\end{eqnarray}
We define `pure condensate' wave functions as
\begin{eqnarray}
|N, \vec{n} \rangle &=& \frac{1}{\cal N}( n_x a_x^{{\dagger}} +
n_y a_y^{{\dagger}} + n_z a_z^{{\dagger}} )^N |0\rangle
\label{purecondensate}
\end{eqnarray}
that minimize the $U_2$ interaction energy at the Gross-Pitaevskii
(mean-field) level in each given well \cite{Castin00}. Here ${\cal
N}=[ 2 (N-1)! ]^{1/2}$ is a normalization factor, which is
calculated in Appendix \ref{AppendixD}.

 Now we can construct states as
\begin{eqnarray}
|\psi \rangle_N &=& \int_{\bf{n}} \psi(\vec{n}) |N, \vec{n}
\rangle, \label{Nbigwavefunction}
\end{eqnarray}
where  $\int_{{\bf n}}$ stands for $\int d {\bf n}/(4\pi)$.

Condition (\ref{SNeven}) corresponds to the symmetry of the states
(\ref{purecondensate})
\begin{eqnarray}
|N, -\vec{n} \rangle = (-)^N |N, \vec{n} \rangle.
\end{eqnarray}
Hence, we need to consider only wave functions that satisfy
$\psi(-\vec{n})=(-)^N \psi(\vec{n}).$

Now we can consider how a spin rotation operator acts on the wave
function $\psi({\bf n})$:
\begin{eqnarray}
e^{-i \theta_\alpha S_\alpha}  |\psi \rangle_N
 = \int_{\bf n} \psi({\bf n}) e^{-i
\theta_\alpha S_\alpha}|N, {\bf n} \rangle  = \int_{{\bf n}'}
\psi( e^{i \theta_\alpha S_\alpha }{\bf n }')|N, {\bf n}' \rangle
= \nonumber \\
 \int_{{\bf n}'} \psi( n_{\gamma} + \epsilon_{\alpha\beta\gamma}
\theta_\alpha n'_{\beta}))|N, {\bf n}' \rangle.
\end{eqnarray}
Expanding the last expression for small $\theta$ we find
\begin{eqnarray}
L_{\alpha} \psi = - i \epsilon_{\alpha\beta\gamma} n_\beta
\frac{\partial}{\partial n_\gamma} \psi, \label{Sn}
\end{eqnarray}
where we used $L_\alpha$ rather than $S_\alpha$ to show that it
acts on the wave function $\psi$. Therefore, operator ${\bf L}$ is
an angular momentum operator for ${\bf n}$. If we want to
construct any spin state, we should take $\psi({\bf n})$ to be
usual spherical harmonic. We note that the $S=0$ result in
\cite{Castin00} is just a special case of our general statement.
The most general form of the state in  a well can be expanded as
$$
\psi({\bf n})=\sum_{|m|\leq L,} c_{L,m} Y_{L m}({\bf n}),
$$
where $L$ satisfies conditions~(\ref{SNeven})-(\ref{SlessN}).

At this point, what we have done is valid for all $N$, not only
big ones. This representation is particularly suitable for $N\gg
1$ since in this limit  states that correspond to different ${\bf
n}$'s are orthogonal to each other(see Appendix \ref{AppendixD})
\begin{eqnarray}
 \langle N, {\bf n}_1 | N, {\bf n}_2 \rangle = \delta_N ({\bf n}_1 - {\bf
n}_2) \label{largeN}
\end{eqnarray}

The delta function is defined from the condition
\begin{eqnarray}
\int_{{\bf n}_1} \int_{{\bf n}_2} f^1_N({\bf n}_1) f^2_N({\bf
n}_2) \delta_N ({\bf n}_1 - {\bf n}_2) = \int_{{\bf n}} f^1_N({\bf
n}) f^2_N({\bf n})
\end{eqnarray}
for the functions that satisfy $f_N(-{\bf n})= (-)^N f_N({\bf
n})$.

We show in the Appendix \ref{AppendixD} that
\begin{eqnarray}
a_\alpha^{{\dagger}} |N, {\bf n} \rangle &=& (N+1)^{1/2} n_\alpha
|N+1, {\bf n} \rangle,
\nonumber\\
a_\alpha |N, {\bf n} \rangle &=& N^{1/2} n_\alpha |N-1, {\bf n}
\rangle \label{create-annih}
\end{eqnarray}
after projecting into the `pure condensate' wave functions.

 This allows us to represent~(\ref{create-annih}) as the product of two
 operators, which act in different spaces. For each trap we define  the particle creation and annihilation
operators that  change the number of particles $N$ but not the
direction of ${\bf n}$\footnote{ The operator $b$ applied directly
to $|\psi\rangle_N $ takes us outside of the physical Hilbert
space since it does not change the symmetry of the $\psi({\bf n})$
wave function simultaneously with changing the number of particles
by one. The physical Hamiltonian, however, will always have this
operator $b$ in combination with some odd function of ${\bf n}$
and will preserve the physical Hilbert space.}
\begin{eqnarray}
b_i^{{\dagger}} |N_i, {\bf n}_i \rangle &=& (N_i+1)^{1/2} |N_i+1,
{\bf n}_i \rangle,
\nonumber\\
b_i |N_i, {\bf n}_i \rangle &=& N_i^{1/2} |N_i-1, {\bf n}_i
\rangle.
\end{eqnarray}
The number of particles in each trap may  be expressed using $b$
operators as
\begin{eqnarray}
N_i = b_i^{{\dagger}} b_i.
\end{eqnarray}

Hamiltonian (\ref{OriginalHamiltonian}) can now be represented as

\begin{eqnarray}
{\cal H}=-t\sum_{\langle i j\rangle,\sigma}{\bf n}_i {\bf n}_j\,
(b_i^{{\dagger}} b_j + b_j^{\dagger} b_i ) - \mu \sum_{i}\hat N_i+
\frac{U_0}{2}\sum_{i}\hat N_i(\hat N_i-1) + \frac{U_2}{2}
\sum_{i}(\vec L_i^2-2 \hat N_i ), \label{HamiltonianRigidRotor}
\end{eqnarray}
where
\begin{eqnarray}
{\bf L}_i = -i\, {\bf n}_i \times \frac{\partial}{\partial {\bf
n}_i}.
\end{eqnarray}

Now, if we are in the Mott insulating phase, we can easily derive
the effective Hamiltonian for $\psi({\bf n})$. Using the second
order perturbation theory, we find the effective Hamiltonian on
the sphere to be
\begin{eqnarray}
{\cal H} = \frac{U_2}{2} \sum_i \vec{L}_i^2 -\frac{2N^2t^2}{U_0}
\sum_{ij}n_{ia}n_{ib}n_{ja}n_{jb}, \nonumber \\
\psi({\bf n})=(-1)^N\psi({\bf -n}).
\label{NbigEffectiveHamiltonian}
\end{eqnarray}
We note that this Hamiltonian corresponds to a lattice of quantum
rotors that interact via quadrupolar moments.

\subsection{Mean Field Solution}
Now we can find a mean field ground state of
(\ref{NbigEffectiveHamiltonian}). We consider the case when the
quadrupolar interaction term is much bigger than the kinetic term.
We show that in this case the ground state is a  uniaxial nematic
and find its energy. Comparing the energy of this state to that of
a singlet, we estimate the phase boundary for nematic-singlet
transition  for even $N$.

Our general mean field anzats has the form
\begin{eqnarray}
|\Psi>=\prod_i\Psi(\theta_i,\varphi_i), \int_{\bf
n}}{|\Psi(\theta,\varphi)|^2 =1.
\label{NbigvariationalWavefunction}
\end{eqnarray}

Expectation value of (\ref{NbigEffectiveHamiltonian}) per well
over the wave function (\ref{NbigvariationalWavefunction}) equals
$$
 \frac{U_2}{2}<\vec{L}_i^2> - J <n_\alpha n_\beta><n_\beta n_\alpha>,
$$
where
\begin{eqnarray}
J= \frac{z N^2 t^2}{U_0} ,
\end{eqnarray}
\begin{eqnarray}
\vec{L}_i^2=-\frac{1}{\sin\theta}\frac{\partial}{\partial\theta}
(\sin\theta
\frac{\partial}{\partial\theta})-\frac{1}{\sin^2\theta}
\frac{\partial^2}{\partial\varphi^2}.
\end{eqnarray}
As in the case of $N=2,$ all of the states that can be transformed
into each other by global  rotation have the same energy.
Therefore, we can impose three additional conditions. The best
choice is to require symmetric real matrix $<n_i n_j>$ to be
diagonal and to choose $<n_z^2>$ to be the biggest eigenvalue. In
such a gauge interaction, the term becomes
$-J(<n_x^2>^2+<n_y^2>^2+<n_z^2>^2).$ Since we have the extra
constraint $<n_x^2>+<n_y^2>+<n_z^2>=1,$ it is now obvious that
interaction energy is minimized when
$$
<n_z^2>=\cos^2\theta\rightarrow 1.
$$
However, states with $\sin^2\theta \rightarrow 0$ have higher
angular moments, and the ground state is determined by the
competition of these two factors. We write mean field
Gross-Pitaevskii equations to determine the ground state
$$
\{-\frac{U_2}{2}\frac{1}{\sin\theta}\frac{\partial}{\partial\theta}
(\sin\theta
\frac{\partial}{\partial\theta})-\frac{1}{\sin^2\theta}
\frac{\partial^2}{\partial\varphi^2}
$$
\begin{equation}
-2J(n_x^2<n_x^2>+n_y^2<n_y^2>+n_z^2<n_z^2>)\}\Psi(\theta,\varphi)=\lambda
\Psi(\theta,\varphi), \label{GP}
\end{equation}
where $\lambda $ is Lagrange multiplier. Now we consider the  case
$J \gg U_2$. In this case interaction energy dominates and we
expect $\sin^2\theta \rightarrow 0$ and, therefore, wave function
becomes localized  near $z$ and $-z$ directions. We can solve the
problem by expanding only near $\theta=0$ and then taking an
(anti)symmetric combination to satisfy
(\ref{NbigEffectiveHamiltonian}). It is obvious that if we expand
the kinetic part of (\ref{GP}) up to the first nonvanishing order
in $\theta,$ then we will get a two dimensional Laplace operator
$\Delta_{n_x,n_y},$ and our problem becomes equivalent to a
harmonic oscillator. Effective parameters are expressed as
\begin{equation}
 m = \frac1{U_2} ,\,\, \omega_{x,y}^2= 4 J U_2 <n_z^2 - n_{x,y}^2>.
 \label{Nbigeffectiveparameters}
\end{equation}

Since we have already neglected higher order terms in $<\theta^2>$
while obtaining a harmonic hamiltonian from (\ref{GP}), with the
same accuracy we can set $\omega_x=\omega_y=\sqrt{4 J U_2}$ in
(\ref{Nbigeffectiveparameters}), i.e. the ground state is a  {\it
uniaxial} nematic. Since we know wave functions, we can calculate
the expectation value of energy. We will use the fact that for a
harmonic oscillator the expectation value of  kinetic energy is
the same as that of potential energy. Energy of the ground state
becomes
$$
E=\frac38(\omega_x+\omega_y) - J{<n_z^2>}.
$$
 Quantum fluctuations of the direction of $ \bf n$ equals
\begin{equation}
<n_{x}^2>=<n_{y}^2>=\frac{U_2}{2\omega}=\frac14\sqrt{\frac{U_2}{J}},
\label{nfluctuation}
\end{equation}
 and expectation value of the ground state energy is
\begin{equation}
E=\frac34\omega+\frac{J U_2}{\omega}-J = 2 \sqrt{U_2 J} - J .
\label{Nbigenergy}
\end{equation}
Symmetrization  or antisymmetrization of the wave function
introduces exponentially small shifts in energy, so in limit $J\gg
U_2$ energy doesn't feel the parity of $N$. Though in this
subsection we explicitly  started from variational anzats
(\ref{NbigvariationalWavefunction}), now we can justify it in the
limit $J\gg U_2$ since  in this case quantum fluctuations of the
direction of $\bf{\hat{n}}$ are small and given by
(\ref{nfluctuation}).

 From section \ref{Nodd} we know that for small nonzero $t$
there is a uniaxial nematic state for odd $N$. Since in the
opposite limit there is also a nematic state, we expect that for
all $N \gg 1, N=2n+1$ the insulating state will be nematic. For
the case of even $N$, there is always a singlet state in which
mean-field energy equals $-J/3$. Comparing this energy with
(\ref{Nbigenergy}), we can estimate the first order transition
point as $U_2=J/9.$ At this point
$$<n_{x}^2>=1/12,$$
so we expect our expansion to be valid.

\section{Global phase diagram}
\label{GlobalPhase diagram}

In the earlier sections we have established spin structure of
insulating phases of  $S=1$ bosons in the optical lattice in
various limits.
% It is easy to see that all of our results fit
%within a simple paradigm: {\it Insulating phases with odd number
%of particles per site are always nematic, and insulating phases
%with even filling factors are spin singlet for small tunnelling
%and spin nematic for larger tunnelling. Transitions between
%singlet and nematic phases are first order with a narrow region
%around the transition in which both phases are locally stable}.
Here we summarize our arguments and discuss
implications of our results for
the global phase diagram.

\subsection{Two and three dimensional lattices}

In two and three dimensional lattices, insulating states with one
atom per site are nematic as long as the perturbation theory
approach in $t/U_0$ remains valid, as was shown in section
\ref{Nodd}. For an arbitrary odd number of particles per site,
$N$, and in the limit of small tunnelling $(Nt)^2/U_0<<U_2$, the
nematic order in the ground state was also established in section
\ref{Nodd}. For large, odd $N$, the nematic order in the ground
state can be proven when $(Nt)^2/U_0$ becomes larger than $U_2$
(but still smaller than $U_0$), as was demonstrated in section
\ref{Nbig}. It is also natural to expect that the superfluid polar
phase develops from the nematic insulator (both states break spin
rotational symmetry without breaking the time reversal symmetry),
so we expect the nematic order even when $Nt/U_0$ is not small and
the system is close to the superfluid-insulator transition. In all
cases we find that insulating phases with an odd number of
particles per site are nematic.

 In the case of two
particles per site, the results of section \ref{N2} establish that
for small enough $U_2/U_0$ there is a first order transition
between the spin singlet phase (for small $t$) and the spin
nematic phase (for larger $t$) at $zt_c^2/U_0U_2=0.5$ ($z$ is the
coordination number of the lattice). Analogously, for large, even
$N$, results of section \ref{Nbig} show that the singlet
insulating ground state goes into spin nematic at
$zN^2t_c^2/U_0U_2=9$. Since for small enough $U_2/U_0$ we expect
nematic spin order close to the SI transition into the polar
superfluid phase, we propose that in this case insulating phases
with an even number of particles per site are either singlet or
nematic with the first order transition at some critical value of
tunnelling $t_c$.

In all of our earlier discussions, we assumed that Mott insulating
lobes for even fillings are big enough to have the transition into
the nematic phase before superfluidity sets in. This assumption is
controlled by the smallness of the ratio $U_2/U_0.$ Here we will
discuss the superfluid - insulator phase boundaries and estimate
how small $U_2/U_0$ should be for the singlet-nematic transition
to lie inside the Mott phase.

\begin{figure}
\psfig{file=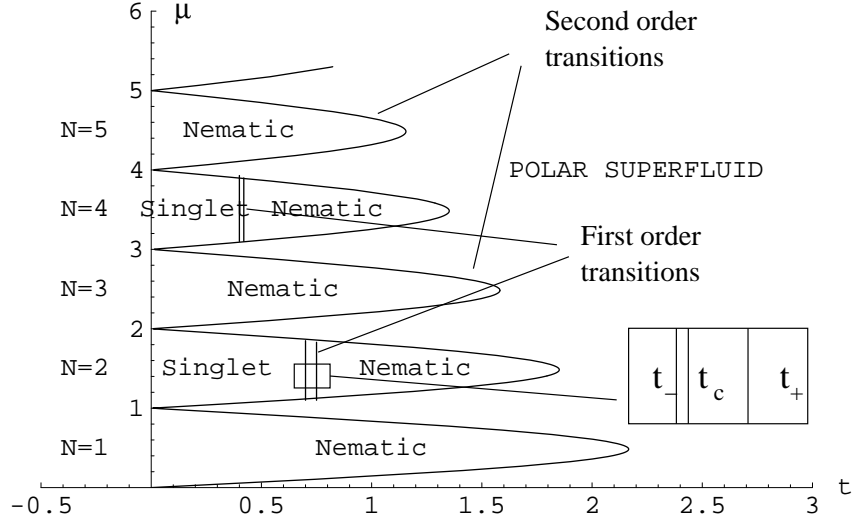} \caption{\label{phasediag} Global phase
diagram for $S=1$ bosons in 2d and 3d optical lattice, for
$zU_2/U_0<0.1$. Mean-field analysis of the superfluid to insulator
transition was done in \cite{japan}. In this work we concentrated
on discussing the spin structure of the insulating lobes.
Singlet-nematic first order phase transition for $N=2$ takes place
for $zt^2/U_0U_2=0.5$ ($z$ is the coordination number of the
lattice). For large, even $N$ singlet-nematic  phase transition
occurs at $zN^2t^2/U_0U_2=9$. $t_c$ marks the actual first order
phase transition and $t_-$ and $t_+$ are the limits of
metastability. Note that the system may also have fragmented
superfluid phases for small $t$ that are not shown here
\cite{DemlerZhou}. }
\end{figure}

\begin{figure}
\psfig{file=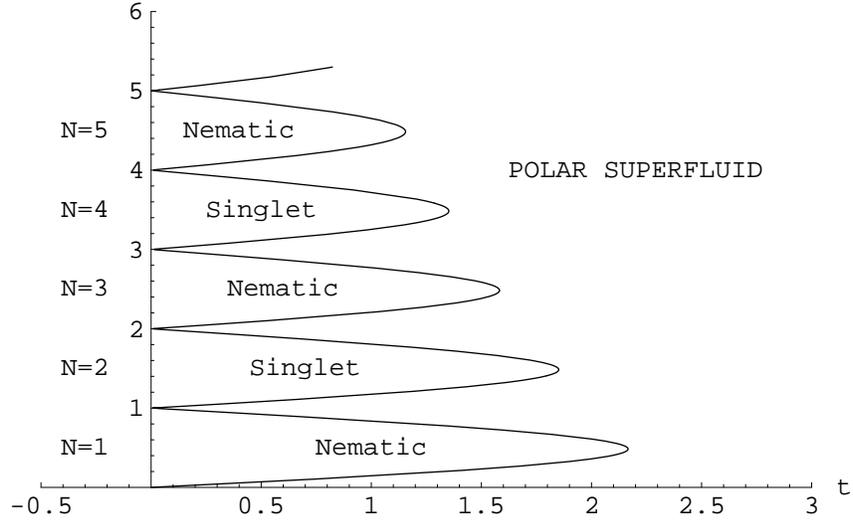} \caption{\label{phasediag2} Global
phase diagram for $S=1$ bosons in 2d and 3d optical lattice, for
$zU_2/U_0>0.1$. Superfluid-Insulator transition for even filling
factors takes place before singlet-nematic transition. }
\end{figure}

Assuming transition from the spin-singlet insulating phase, the
mean-field calculation of the superfluid-insulator phase boundary
was given in \cite{japan}. Analysis presented in this paper shows
that the critical value of tunnelling, after which the Mott phase
doesn't exist, is given by
 $$
 \frac{U_0+2 U_2}{z t_{SI}}=\frac13(2N+3+2\sqrt{N^2+3N}).
 $$
We will use this critical value $t_{SI}$ as an estimate of the
superfluid-insulator transition. For $N=2$, singlet-nematic phase
transition takes place at $zt_c^2/U_0U_2=0.5.$ The condition
$t_c<t_{SI}$ for $N=2$ is satisfied, if
$$
\frac{zU_2}{U_0} < 0.1.
$$
For the case  $N>>1$ the requirement of $t_c<t_{SI}$  becomes even more
restrictive, namely
$$
\frac{zU_2}{U_0} < 0.01.
$$
One can see that depending on the exact value of $z U_2/U_0,$
there are different possibilities for Mott lobes with an even
number of particles. When $zU_2/U_0<0.01,$ all insulating phases
with even filling factors are spin singlet for small tunnelling
and spin nematic for larger tunnelling. For $0.01<zU_2/U_0<0.1,$
insulating phases with small, even filling factors have both
singlet and nematic regimes, but insulating states with
sufficiently large even fillings have only the singlet phase.
Finally, for $0.1<zU_2/U_0$, all insulating phases with even
filling factors are in the spin singlet state. In Figs.
\ref{phasediag} and \ref{phasediag2} we combine these results with
the schematic representation of the SI transitions to obtain the
global phase diagram.

\subsection{One dimensional lattices}

For one dimensional lattices we established that when $N^2t^2/U_0
<< U_2$ the system will be in a uniform singlet phase for even
fillings and in a dimerized singlet phase for odd fillings (when
there is only one atom per site the dimerized phase has been
verified in the regime $t << U_0$). The nature of magnetic order
close to the tips of the insulating lobes (when the perturbation
theory in $t$ is not applicable) is less clear. However, we expect
that the phase diagram for the one dimensional lattice is
qualitatively similar to two and three dimensional cases with one
important difference: instead of the nematic phase, we have
dimerized singlet states. This will be discussed in future
publications.

\section{Detection of spin order in insulating phases}
\label{experiment}

Now we discuss two approaches to  detection of  the
singlet-nematic phase transition for $S=1$ bosons in an optical
lattice. One  way of detecting such a transition has already been
noted in section \ref{magneticfieldeffects}, where we proposed to
introduce an easy plane for nematic order by applying a small
magnetic field, then releasing the trap and measuring the number
of particles of different spin components. Spatial separation of
different spin components can be achieved by applying magnetic
field gradients during the free fall of the atoms. For the case of
$N=2$, with a small magnetic field applied in the $z$ direction,
expectation values of $n(S_z=0)$ and $n(S_z=1)=n(S_z=-1)$ have
been calculated and are shown in Fig. \ref{n0n1}. Since the phase
transition is of the first order, there is a sharp change which
can be measured experimentally. We note that $N=2$ case also have
particular experimental advantage over other filling factors due
to the absence of three particle losses and the least restrictive
condition on $U_2/U_0$ for observation of singlet-nematic
transition.
%Experiments  of this kind can be performed for
%different directions of initial magnetic field and magnetic field
%gradients, and signature of nematic phase is that for any
%direction of initial magnetic field and field gradient $\vec d$
%number of particles with $\pm 1$ spin component along $d$ axis
%coincide: $n(S_{\vec{d}}=1)$ = $n(S_{\vec{d}}=-1).$ This condition
%is satisfied since in nematic phase time reversal symmetry in not
%broken and spin operators have zero expectation values, $<\vec
%S>=0.$

\begin{figure}
\psfig{file=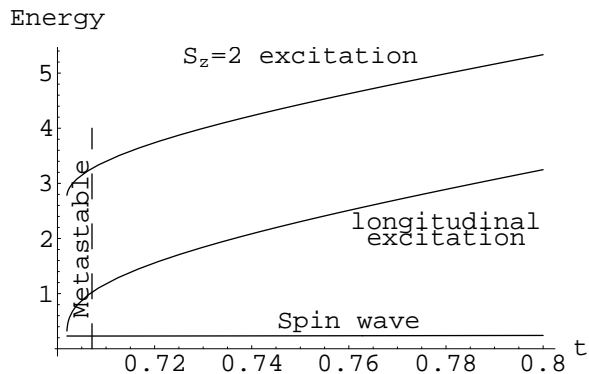} \caption{\label{bragg2} Positions of
Bragg peaks for $|{\bf q}|=0.02,z=6$ in the nematic
phase with $N=2$ atoms per site. Energy is measured in units
of $U_2,$ and $t$ is measured in units of $\sqrt{\frac{U_0
U_2}{z}}$}
\end{figure}

The second approach  to experimental detection of singlet and
nematic insulating phases relies on the measurement of  excitation
spectra. As discussed in sections \ref{spinsingletexcitations} and
\ref{nematicspinwaves}, the singlet phase has a nonzero gap to all
excitations, whereas the nematic phase has  gapless spin wave
excitations. To measure the excitation spectra, we propose using
Bragg spectroscopy, which was used successfully to identify
sound-like Bogoliubov excitations in condensates of spinless
particles \cite{MITbraggscattering}.
\begin{figure}
\psfig{file=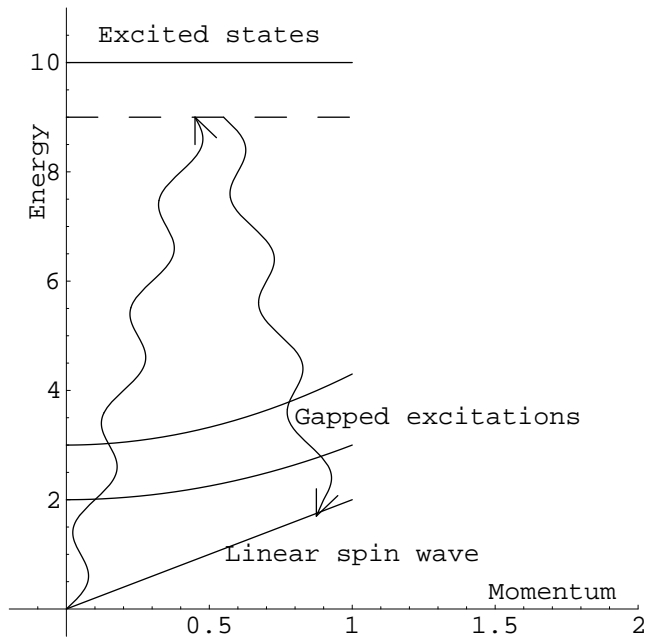} \caption{\label{bragg}Probing dispersion
relation using Bragg scattering. }
\end{figure}
In such experiments the optical lattice should be illuminated by
two laser beams with wave vectors $\bf k_1$ and $\bf k_2$ and a
frequency difference $\omega,$ which is much smaller than their
detuning from an atomic resonance. The intersecting beams create a
periodic, travelling intensity modulation that creates external
potential due to ac Stark effect of the form
$V_{\alpha\beta}\cos{({\bf q r }-\omega t)}.$ Here we introduce
spin indices for the external potential since the ac Stark effect
may introduce mixing between different $S_z$ components. A
response of the system to such potential may be calculated using
Fermi's golden rule. Interaction is expressed in second-quantized
notations such as $V_{\alpha \beta}/2(\hat \rho_{\alpha
\beta}^{\dagger}({\bf q })e^{-i \omega t}+\hat \rho_{\alpha
\beta}^{\dagger}({\bf -q }) e^{i \omega t} ) ,$ where $\hat
\rho_{\alpha \beta}^{\dagger}=\sum_k a^{\dagger}_{\alpha k+q}
a_{\beta k}.$ The scattering rate is given by
$$
\frac{2 \pi}{\hbar}\sum_f |\langle f|V_{\alpha \beta}\hat
\rho^{\dagger}_{\alpha \beta} |g\rangle|^2 \delta(\hbar \omega
-E_f+E_g).
$$
If the resonance state is far detuned from the excited states,
then $V_{\alpha \beta}$ has the form $V\delta_{\alpha \beta},$ and
couples only to the total number of particles in each well and
doesn't feel internal spin structure. Low lying excitations in
insulating phases don't change the number of particles on
individual sites, so $V\delta_{\alpha \beta}$ interaction won't
produce any Bragg peaks for low lying excitations. Therefore, it
is necessary for detection that $V_{\alpha \beta}$ deviate from
$V\delta_{\alpha \beta},$ which can be achieved by making detuning
comparable to level spacing of fine and hyperfine components. From
section \ref{nematicspinwaves} we know that for $N=2$, nematic
spin wave excitations correspond to $S_z=\pm 1,$ longitudinal
excitation corresponds to $S_z=0$, and there are also gapped
excitations with $S_z=\pm 2.$ Since the nematic state has $S_z=0,$
it is necessary to have nonzero $V_{0,\pm 1}$, $V_{0,0},$ and
$V_{\pm 1,\mp 1} $ to observe each kind of these excitations. In
Fig. \ref{bragg2} we show dependence of $N=2$ peak positions on
$t$ for fixed $\bf q.$

Finally, we consider the effect of inhomogeneous trapping
potential. When this local trapping potential $\varphi_i$ varies
smoothly from site to site, it is not the chemical potential
$\mu_i$ which is fixed across the trap, but the sum $\varphi_i+
\mu_i.$ Therefore, if $\varphi_i$ varies considerably, we will
have insulating regions with different occupation numbers as well
as regions with superfluid order, all in the same trap, as was
discussed in \cite{Jaksch98} for the case of spinless atoms.
Therefore, Bragg scattering experiments for fixed ${\bf q }$ will
exhibit resonances coming from the regions of the lattice at
different filling factors. Relative intensity of these resonances
will be determined by the relative number of particles in each
region. Interpretation of Bragg experiments will be easier if the
trapping potential is not harmonic, but has sharp borders, so the
whole system has essentially the same density.

\section{Conclusions}

In summary, we have considered Mott insulating phases of spin-1
atoms with antiferromagnetic interactions in optical lattices.
%Depending on parameters of optical lattice, Mott phase exhibits
%singlet or nematic ground state.
In the experimentally interesting limit $U_2\ll U_0$, and deep
inside the Mott phases $N t \ll U_0$ ($N$ is the filling factor),
we performed detailed calculations for  the following cases: i)
odd number of particles per site and $(N t)^2/U_0 \ll U_2$; ii)
two particles per site and an arbitrary ratio of $t^2/U_0$ and
$U_2$; iii) large number of particles per site $N \gg 1$ with an
arbitrary ratio of $(Nt)^2/U_0$ and $U_2$. Based on this analysis
we argued that in two and three dimensional lattices insulating
phases with an odd number of particles per site are always
nematic. For an even number of particles per site, there is either
a spin singlet phase or a first order phase transition between
spin singlet and nematic phases controlled by the depth of the
optical lattice. The resulting global phase diagrams are shown in
Fig. \ref{phasediag2} and \ref{phasediag}. We have considered
excitations for singlet and nematic phases and have reviewed the
effects of small magnetic field. For one dimensional lattices we
have found dimerized singlet phases for insulating states with odd
fillings. We also discussed different experimental techniques to
identify the proposed phases.

We thank E. Altman, D. Haldane, W. Hofstetter, D. Podolsky, S.
Sachdev, A. Sorensen, D.W. Wang, and F. Zhou for useful
discussions. This work was partially supported by the NSF Career
Award DMR-0132874, PHY-0134776, and by the Sloan and Packard
Foundations.  When this work neared completion, we learned that
similar results have been obtained by M. Snoek and F.
Zhou\cite{snoekzhou}.

\appendix

\section{Derivation of the Effective Magnetic Hamiltonian for An
Insulating State with Odd Number of Atoms} \label{AppendixA}

To be able to derive $J_0, J_1, J_2$ dependance on $n$, we should
know how to write down explicitly, in terms of creation and
annihilation operators, all of the states that we are interested
in.
 To do this we introduce singlet pair creation operator(summation over
 repeated indices is presumed over $\{x,y,z\}$):

 $$
    \theta^{\dagger}= a^{\dagger}_p a^{\dagger}_p=(a^{\dagger}_0)^2 - 2 a^{\dagger}_{-}a^{\dagger}_{+},
 $$
which has the following commutation relations:

 $$
    [a_p,\theta^{\dagger}]= 2 a^{\dagger}_p, [a_0,\theta^{\dagger}]=2 a^{\dagger}_0, [a_\pm,\theta^{\dagger}]=- 2
    a^{\dagger}_\mp,
 $$
\begin{equation}
[ a^{\dagger},\theta^{\dagger}]=0,[
S_{a},\theta^{\dagger}]=0,[\vec S^2,\theta^{\dagger}]=0.
\label{Sthetacommutator}
\end{equation}
Since $\theta^{\dagger}$ commutes with spin operators, it doesn't
change total spin  and spin components but just adds 2 electrons.
Therefore, we can construct unnormalized spin states for arbitrary
$n$ in the following way: first, write down a state with necessary
spin for a small number of particles; second, apply
$\theta^{\dagger}$ as many  times as needed to get the desired
number of particles.

 Using this procedure we obtain that states
$$a_+^{\dagger} (\theta^{\dagger})^n |0>, a_{-}^{\dagger} (\theta^{\dagger})^n |0>,a_0^{\dagger}
(\theta^{\dagger})^n |0>$$ belong to $S=1$ low energy subspace and
are orthogonal since they are different eigenvectors of $S_{z}$.
Making orthogonal transformation that leads to
$\{a_x^{\dagger},a_y^{\dagger},a_z^{\dagger}\}$ basis, we can
write three orthonormal states with $S=1$ as
$$
|S=1,p,2n+1>=\frac1{\sqrt{f(n;1)}}a_p^{\dagger}
(\theta^{\dagger})^n |0>,
$$
where normalization factor
$$
 f(n;s)=s!n!2^n\frac{(2n+2s+1)!!}{(2s+1)!!}
$$
was calculated in~\cite{HoYip}. In our calculation  later we will
need more general normalization factors, so first we will derive
the way to normalize our spin states.

\subsection{Normalization of the states}
 We will be interested in normalization of  the states
$$
|a,b,n>=a^{\dagger}_a a^{\dagger}_b (\theta^{\dagger})^n |0>,
$$
and calculation of
$$
f(a,b,p,q,n)=<a,b,n|p,q,n>= <0|(\theta)^n a_a a_b a^{\dagger}_p
a^{\dagger}_q (\theta^{\dagger})^n |0>,
$$
where $a,b,p,q \in \{x,y,z\}$. Let's consider a coherent state
$$
e^{a^{\dagger}_x x_1+a^{\dagger}_y x_2+a^{\dagger}_z x_3}|0>.
$$
We observe that this state is a linear combination of Fock states,
with the coefficients being polynomials in $\{x_1,x_2,x_3\}$. To
extract the weight of the state $|a,b,n>$, we need to calculate
the quantity
$$
|a,b,n>=[T^n_{a,b}(x) e^{a^{\dagger}_x x_1+a^{\dagger}_y
x_2+a^{\dagger}_z x_3}|0>]_{\bf x=0},
$$
where
$$
T^n_{a,b}(x)=\nabla_{x_a}\nabla_{x_b}(\Delta_x)^n.
$$
We  use  the normalization condition for coherent states(see
e.g.~\cite{negele})
$$
<0|e^{a_x y_1+a_y y_2+a_z y_3}e^{a^{\dagger}_x x_1+a^{\dagger}_y
x_2+a^{\dagger}_z x_3}|0>=e^{x_1 y_1+x_2 y_2+x_3 y_3}
$$
to calculate
$$
[T^n_{a,b}(x)T^n_{p,q}(y)e^{x_1 y_1+x_2 y_2+x_3 y_3}]|_{\bf
x=0,y=0}=
$$
$$
  T^n_{a,b}(x)T^n_{p,q}(y)\frac{(x_1 y_1+x_2 y_2+x_3
y_3)^{2n+2}}{(2n+2)!}.
$$
We can expand $T^n_{a,b}(x)T^n_{p,q}(y)$ using the extended Newton
binomial formula :
$$
T^n_{a,b}(x)T^n_{p,q}(y)=\sum_{n_1+n_2+n_3=n}\frac{n!}{n_1!n_2!n_3!}\nabla_{x_1}^{2
n_1}\nabla_{x_2}^{2 n_2} \nabla_{x_3}^{2
n_3}\nabla_{x_a}\nabla_{x_b} \times
$$
$$
\sum_{m_1+m_2+m_3=n}\frac{n!}{m_1!m_2!m_3!}\nabla_{y_1}^{2
m_1}\nabla_{y_2}^{2 m_2} \nabla_{y_3}^{2
m_3}\nabla_{y_p}\nabla_{y_q}.
$$
Let's first consider the case when  sets $\{a,b\}$ and $\{p,q\}$
coincide. We have two essentially different cases: $a=b$ and $a\ne
b$. Without loss of generality, suppose for first case
$a=b=p=q=x$. Then,
$$
<a,a,n|a,a,n>=\sum_{n_1+n_2+n_3=n,l_1+l_2+l_3=2n+2}(\frac{n!}{n_1!n_2!n_3!})^2\frac{1}{l_1!l_2!l_3!}
\times
$$
$$
 (\nabla_{x_1}\nabla_{y_1})^{2 n_1+2} (\nabla_{x_2}
\nabla_{y_2})^{2 n_2}(\nabla_{x_3}\nabla_{y_3})^{2 n_3}(x_1
y_1)^{l_1}(x_2 y_2)^{l_2}(x_3 y_3)^{l_3} =
$$
$$
\sum_{n_1+n_2+n_3=n}(\frac{n!}{n_1!n_2!n_3!})^2{(2n_1+2)!(2n_2)!(2n_3)!}.
$$
This double sum can be calculated; the answer is
$$
f(x,x,x,x,n)=\frac{2}{15}(3+2n)(5+3n)(2n+1)!.
$$
For the case of different indices, the normalization is
$$
f(x,y,x,y,n)=\sum_{n_1+n_2+n_3=n,l_1+l_2+l_3=2n+2}(\frac{n!}{n_1!n_2!n_3!})^2\frac{1}{l_1!l_2!l_3!}
\times
$$
$$
 (\nabla_{x_1}\nabla_{y_1})^{2 n_1+1} (\nabla_{x_2}
\nabla_{y_2})^{2 n_2+1}(\nabla_{x_3}\nabla_{y_3})^{2 n_3}(x_1
y_1)^{l_1}(x_2 y_2)^{l_2}(x_3 y_3)^{l_3} =
$$
$$
\sum_{n_1+n_2+n_3=n}(\frac{n!}{n_1!n_2!n_3!})^2{(2n_1+1)!(2n_2+1)!(2n_3)!}=
$$
$$
\frac{1}{15}(3+2n)(5+2n)(2n+1)!.
$$
Now let's consider the case when $\{a,b\}$ and $\{p,q\}$ don't
coincide. There are 4  essentially different cases:
$$
(x,y,x,z),(x,x,y,z),(x,x,x,y),(x,x,y,y).
$$
All other normalizations can be obtained from these by proper
permutation of indices. In the first three cases, overlap is zero
since(say, for the first case) a nonzero value comes from the term
that satisfies the conditions
 $$2 n_1+1=2 m_1+1, 2 n_2+1=2 m_2,2 n_3=2 m_3+1,$$
which doesn't have integer solutions. However, in the fourth case,
overlap of the  states is not zero: it comes from the terms
obeying
$$
n_1+1=m_1, n_2=m_2+1, n_3=m_3.
$$
We can calculate this quantity analogously to the previous
calculation, but we can use another trick:
$$
<0|(\theta)^{n+1}(\theta^{\dagger})^{n+1}|0>=3 <0|(\theta)^{n}a_x
a_x a^{\dagger}_x a^{\dagger}_x(\theta^{\dagger})^{n}|0> +
6<0|(\theta)^{n}a_x a_x a^{\dagger}_y
a^{\dagger}_y(\theta^{\dagger})^{n}|0>.
$$
Therefore,
$$
f(x,x,y,y,n)=\frac16((2n+3)! - 3
\frac{2}{15}(3+2n)(5+3n)(2n+1)!)=\frac{2}{15}n(3+2n)(2n+1)!.
$$
Since sometimes it will be more convenient for us to work in $\{
+,-,0 \}$ basis, let's also write down all nonzero overlaps in
this basis (up to trivial permutations):
$$
f(\pm,\pm,\pm,\pm,n)=<0|(\theta)^{n}a_\pm a_\pm a^{\dagger}_\pm
a^{\dagger}_\pm(\theta^{\dagger})^{n}|0>=\frac{2}{15}(5+2n)(3+2n)(2n+1)!,
$$
$$
f(0,0,0,0,n)=<0|(\theta)^{n}a_0 a_0 a^{\dagger}_0
a^{\dagger}_0(\theta^{\dagger})^{n}|0>=\frac{2}{15}(5+3n)(3+2n)(2n+1)!,
$$
$$
f(0,\pm,0,\pm,n)=<0|(\theta)^{n}a_\pm a_0 a^{\dagger}_\pm
a^{\dagger}_0(\theta^{\dagger})^{n}|0>=\frac{1}{15}(5+2n)(3+2n)(2n+1)!,
$$
$$
f(+,-,+,-,n)=<0|(\theta)^{n}a_+ a_- a^{\dagger}_+
a^{\dagger}_-(\theta^{\dagger})^{n}|0>=\frac{1}{15}(5+4n)(3+2n)(2n+1)!,
$$
$$
f(+,-,0,0,n)=<0|(\theta)^{n}a_+ a_- a^{\dagger}_0
a^{\dagger}_0(\theta^{\dagger})^{n}|0>=-\frac{2}{15}n(3+2n)(2n+1)!.
$$

\subsection{Calculation of $\epsilon_0$}
 So, first let's calculate the energy  for total spin 0 state. From known
Clebsch-Gordon coefficients, this state is
$$
|S=0,S_z=0>=\frac1{\sqrt3}(|1,1>_i|1,-1>_j-|1,0>_i|1,0>_j+|1,-1>_i|1,1>_j).
$$
Rewriting it via creation operators, we get the normalized state
$$
|S_i+S_j=0>=\frac{1}{\sqrt3
f(n,1)}a^{\dagger}_{i,p}a^{\dagger}_{j,p}(\theta_i^{\dagger})^n(\theta_j^{\dagger})^n|0>.
$$
In the  second order  of perturbation theory, the energy
expectation value is
$$
\sum_{m}-\frac{t^2}{E_m-E_0}|<m|(a^{\dagger}_{i p}a_{j
p}+a^{\dagger}_{j p}a_{i p})|S_i+S_j>|^2.
$$
The intermediate state $|m>$ cannot correspond to $S_i=2,S_j=0$
since in  this case two spins can't add to form  a singlet. There
are always at least two possible  states:
 $$S_i=0,n_i=2n+2,S_j=0,n_j=2n,S_i+S_j=0$$ and $i\leftrightarrow j$.
 The matrix element for each of these states is
 $$
\frac{1}{\sqrt3
f(n,1)}<0|\theta^{n+1}a^{\dagger}_{p}a^{\dagger}_{q}
(\theta^{\dagger})^n|0>_i<0|\theta^n a_p a^{\dagger}_q
(\theta^{\dagger})^n|0>_j\frac{1}{\sqrt{\left(
f(n+1,0)f(n,0)\right)}}=
$$
$$
\frac{1}{\sqrt3 f(n,1)}<0|\theta^{n+1}a^{\dagger}_p a^{\dagger}_q
(\theta^{\dagger})^n|0>_i \delta_{p
q}f(n,1)\frac{1}{\sqrt{\left(f(n+1,0)f(n,0)\right)}}=\sqrt{\left(\frac{f(n+1,0)}{3f(n,0)}\right)}.
$$
Finally, this term gives $$
-\frac{2t^2}{3(U_0-2U_2)}\frac{f(n+1,0)}{f(n,0)}=
-\frac{4t^2(n+1)(2n+3)}{3(U_0-2U_2)}.$$

In general case, $n\ne 0,$  there can also be intermediate states
$$S_i=2,n_i=2n+2,S_j=2,n_j=2n,S_i+S_j=0$$ and $i\leftrightarrow j$
because $S_i=2$ and $S_j=2$ can form a singlet. However, in the
case of $n=0$, such a term doesn't exist but can appear if $a_{j
p}$, acting on $a^{\dagger}_{j q} (\theta_j^{\dagger})^n|0>$,
breaks a singlet part and the result has a mixture of $S=2.$ But
for $n=0$ there is no singlet part, so these intermediate states
are absent. One should also note that even for big $n$ we won't
have terms with higher spins, i.e. $S_i=4,$ because our
perturbation in the Hilbert space of $i-th$ site is a vector and
can have matrix elements only for the  states with spins differing
by $\pm 1.$

 From general considerations we know that for $S_i+S_j=0$ energy has the form
$$
- 2 t^2(\frac{f_1(n)}{U_0 - 2 U_2}+\frac{f_2(n)}{U_0 + 4 U_2}),
$$
where the term with $U_0 - 2 U_2$ in the denominator comes from
the processes of the first kind, and the term with $U_0 + 4  U_2$
in the denominator comes from the processes  of the second kind.
We calculated $f_1(n)$ earlier to be $2(n+1)(2n+3)/3.$ Now, to
find $f_2(n)$, we take the limit $U_2\rightarrow 0.$ In this case
we don't need to know the exact form of states with $S_i=S_j=0$
and $S_i=S_j=2,S_i+S_j=0$ since their energy is the same. We can
take the intermediate state $|m>$ to be
$$
|m>=\frac1{\sqrt{M}}(a^{\dagger}_{i,p}a_{j,p}+a^{\dagger}_{j,p}a_{i,p})a^{\dagger}_{i,q}a^{\dagger}_{j,q}(\theta_i^{\dagger})^n(\theta_j^{\dagger})^n|0>,
$$
where $M$ is normalization of the intermediate state. Then, the
second order energy is
\begin{equation}
-t^2\frac{|<m|m>\frac{\sqrt{M}}{\sqrt3
f(n,1)}|^2}{U_0}=-\frac{t^2}{U_0}\frac{M}{ 3f(n,1)^2}.
\label{Secondorderenergy}
\end{equation}
Using formulas from the previous subsection, we can calculate $M.$
Then, we can write an equation
$$
2(f_1(n)+f_2(n))=\frac{M}{3f(n,1)^2}=2 + \frac65 n(5 + 2 n),
$$
and finally obtain
$$
\epsilon_0 = -\frac{4t^2(n+1)(2n+3)}{3(U_0-2 U_2)}
-\frac{16t^2n(5+2n)}{15( U_0+ 4 U_2)}.
$$

\subsection{Calculation of $\epsilon_1$}
Let's consider the case $S_i+S_j=1.$ There are no intermediate
states $S_i=0,S_j=2$ or $S_i=0,S_j=0$ since these states can only
form $S_i+S_j=2$ and $S_i+S_j=0$, respectively. $S_i=2,S_j=2$ can
add up to form $S_j+S_j=1$, and this is the only contribution.
$S_i+S_j=1$ subspace is three dimensional, and from rotational
invariance we can choose any state we want to calculate the
energy. Let's choose our initial state to be
$S_i=1,S_j=1,S_i+S_j=1,S_{i z }+S_{j z }=0.$ From known
Clebsch-Gordon  coefficients, we can write this state using
creation and annihilation operators as :

$$
|S_i+S_j=1>=\frac{a^{\dagger}_{i,+}a^{\dagger}_{j,-}-a^{\dagger}_{i,-}a^{\dagger}_{j,+}}{\sqrt{2}f(n,1)}(\theta_i^{\dagger})^n(\theta_j^{\dagger})^n|0>.
$$
Normalization of
$$
(a^{\dagger}_{i,+}a_{j,+}+a^{\dagger}_{i,0}a_{j,0}+a^{\dagger}_{i,-}a_{j,-})(a^{\dagger}_{i,+}a^{\dagger}_{j,-}-a^{\dagger}_{i,-}a^{\dagger}_{j,+})|S_i+S_j=1>
$$
equals
$$
2(f(+,0,n)(-2n)^2f(+,0,n-1)+f(+,+,n)(2n)^2f(+,+,n-1))=
$$
$$
\frac{4}{45}n(1 + 2 n)(3 + 2 n)^2(5 + 2n)(2n+1)!^2.
$$
Hence, the energy for this case is
$$
\epsilon_1=-\frac{4 t^2n(5 + 2n)}{5(U_0 + 4 U_2)}.
$$

\subsection{Calculation of $\epsilon_2$}
Lets choose the normalized state for which we will calculate the
second order energy to be the state with total $S_{iz}+S_{jz}=2:$
$$
|S_i+S_j=2>=\frac{1}{
f(n,1)}a^{\dagger}_{i,+}a^{\dagger}_{j,+}(\theta_i^{\dagger})^n(\theta_j^{\dagger})^n|0>.
$$
The intermediate state should also have
$S_i+S_j=2,S_{iz}+S_{jz}=2.$ Such a state can't belong to
$S_i=0,S_j=0$ subspace since this pair of  spins can't add up to
form $S_i+S_j=2.$ There are four possible intermediate states with
$$S_i=2,n_i=2n+2,S_j=0,n_j=2n,$$
$$S_i=0,n_i=2n+2,S_j=2,n_j=2n,$$
 and $i\leftrightarrow j$.
In the first case, matrix element equals
 $$
\frac{1}{f(n,1)}<0|\theta^{n}
a_{+}a_{+}a^{\dagger}_{+}a^{\dagger}_{+}
(\theta^{\dagger})^n|0>_i<0|\theta^n a_+ a^{\dagger}_+
(\theta^{\dagger})^n|0>_j\frac{1}{\sqrt{ f(n,2)f(n,0)}}=
$$
$$
\frac{1}{f(n,1)}f(n,2)f(n,1)\frac{1}{\sqrt{\left(f(n,2)f(n,0)\right)}}=\sqrt{\left(\frac{f(n,2)}{f(n,0)}\right)}.
$$
Finally, this term contributes to the energy
$$
-\frac{2t^2}{(U_0+U_2)}\frac{f(n,2)}{f(n,0)}=
-\frac{4t^2(2n+3)(2n+5)}{15(U_0+U_2)}.$$
 The second process contributes with the same energy denominator
 but with a different dependance on $n$:
 $$
-\frac{16t^2n(n+1)}{15(U_0+U_2)}.
 $$

Now, let's consider the contribution from the
$S_i=2,S_j=2,S_i+S_j=2$ intermediate state. Using the same trick
as at the end of the calculation of $\epsilon_0,$ we only need to
calculate norm of
$$
(a^{\dagger}_{i,+}a_{j,+}+a^{\dagger}_{i,0}a_{j,0}+a^{\dagger}_{i,-}a_{j,-})a^{\dagger}_{i,+}a^{\dagger}_{j,+}(\theta_i^{\dagger})^n(\theta_j^{\dagger})^n|0>.
$$
This quantity equals
$$
f(+,+,n)(f(0,0,n-1)-2(2n+2)f(+,-,0,n-1)+(2n+2)^2 f(+,-,n-1))+
$$
$$
f(+,0,n)(2n)^2f(+,-,n-1) + f(+,-,n)(-2n)^2f(+,+,n-1)=
$$
$$
\frac2{225} (3 + 2 n)^2(5 + 3n)(5 + 6n)(2 n+1)!^2.
$$
Then, taking a limit $U_2\rightarrow0,$ we obtain the
contribution from this process to be
$$
-\frac{28 t^2 n (5 + 2n)}{75(U_0 + 4 U_2)}.
$$
Finally,
$$
\epsilon_2=-\frac{28t^2n(5+2n)}{75(U_0+4U_2)}-\frac{4 (15 + 20 n +
8 n^2)}{15(U_0+U_2)}.
$$

\section{Derivation of the Effective Magnetic Hamiltonian for
the Insulating State with Two Atoms} \label{AppendixB}

To derive the effective Hamiltonian, we should be able to
calculate matrix elements in (\ref{calH}). Since  energy and
matrix elements in each subspace don't depend on $z$ projection of
total spin, we can choose $S_z$ components at our convenience. We
can express any state $|E_1>,...,|E_8>$ using known Clebsch-Gordon
coefficients. For the state from $|E_8>$ with $S_z=0$  , we have
$$
|E_8,S_z> =\sum_{m=-2,...,2}C^{4,0}_{2,m,2,-m}|N_1=2,S_1=2,S_{1
z}=m>|N_2=2,S_2=2,S_{2 z}=-m>.
$$
Now we can write any one of the states $ |N_i,S_i,S_{i z}>$ via
creation and annihilation operators since we know how to express
spin operators via creation and annihilation operators
(\ref{Spinviaa}). Evaluation of $\tilde {e_6} - \tilde{e_8}$ is
quite simple since total spin conservation of the tunnelling term
doesn't allow mixing of this subspaces with any other. Therefore,
as in (\ref{Secondorderenergy}), we just need to calculate the
normalization of the state into which we hop. Using this
procedure, we obtain energies (\ref{epsilontilda68}).

Now let's consider energy in the $|E_1>,|E_2>$ subspace. From
$|E_1>$ we can hop  only into  high energy states
\begin{eqnarray}
|N_1=3,N_2=1,S_1=S_2=1,S_1+S_2=0>, \nonumber \\
|N_1=1,N_2=3,S_1=S_2=1,S_1+S_2=0>, \label{highenergy}
\end{eqnarray}
since in the Hilbert space  of each well spin can change only by
$\pm 1.$ For $N=2$ from $|E_2>$, we can  also tunnel only to these
states since $3$ and $1$ cannot add to form total spin $0,$ and
there is no state $S_1=S_2=3$ which can also add up to total spin
0. Therefore, our exact Hamiltonian in the basis of $|E_1>,|E_2>$
and high energy states (\ref{highenergy}) has the form
$$
\left[\begin{array}{cccc}
  0 & 0 & V_1 & V_1\\
  0 & 6U_2 & V_2& V_2 \\
  V_1 & V_2 & U_0 & 0 \\
  V_1 & V_2 & 0 & U_0
\end{array}\right].
$$
We can diagonalize this hamiltonian in the low energy
$|E_1>,|E_2>$ subspace in the limit $$V_1,V_2 \ll U_0, U_2\ll
U_0.$$
 First, we integrate out high energy levels -- this is done as
 described in \cite{assa}. We use the following matrix identity:
$$
\left[\left( \begin{array}{cc}
   A & B \\
   C & D
 \end{array}\right)^{-1}\right]_{ij}=[(A-B D^{-1} C)^{-1}]_{ij},
$$
where $1\leq i,j \leq2.$

In our case, $D$ has the form $U_0 I_2,$  so it is  easy to
calculate inverse matrix. Finally, our effective Hamiltonian has
the form
$$
\left[\begin{array}{cc}
  0 & 0 \\
  0 & 6 U_2
\end{array}\right]-
\frac2{U_0}\left[\begin{array}{cc}
  V_1^2 & V_1 V_2 \\
  V_1 V_2 & V_2^2
\end{array}\right].
$$
Now we can diagonalize this $2\times2$ matrix; its energy levels
are
$$
3U_2-\frac{V_1^2+V_2^2}{U_0}\pm\sqrt{(3U_2-\frac{V_1^2+V_2^2}{U_0})^2+
12 \frac{U_2 V_1^2}{U_0}}.
$$
Using expressions for all states of interest in Fock basis, we can
calculate
$$
 V_1=-t\sqrt{\frac{10}{3}}, V_2=-t\sqrt{\frac{8}{3}},
$$
which leads to (\ref{epsilon1})-(\ref{epsilon2}).

Now, let's calculate  energy for the $S_i+S_j=2$ subspace. In this
case we can hop to 4 states:
$$
|N_1=3,N_2=1,S_1=S_2=1,S_1+S_2=2>,
$$
$$
|N_1=1,N_2=3,S_1=S_2=1,S_1+S_2=2>,
$$
$$
|N_1=1,N_2=3,S_1=1,S_2=3,S_1+S_2=2>,
$$
$$
|N_1=3,N_2=1,S_1=3,S_2=1,S_1+S_2=2>.
$$
Matrix $A$ has the form(in $E_3, E_4, E_5$ basis)
$$
\left[\begin{array}{ccc}
  3 U_2 & 0 & 0 \\
  0 & 3 U_2 & 0 \\
  0 & 0 & 6 U_2
\end{array}\right],
$$
and matrix B has the form
$$
\left[\begin{array}{cccc}
  V_1 & V_2 & V_3 & 0 \\
  V_2 & V_1 & 0 & V_3 \\
  V_4 & V_4 & V_5 & V_5
\end{array}\right],
$$
and $C=B^T.$ The effective Hamiltonian is
$$
 \left[\begin{array}{ccc}
  3 U_2 & 0 & 0 \\
  0 & 3 U_2 & 0 \\
  0 & 0 & 6 U_2
\end{array}\right]-
\frac1{U_0}
 \left[\begin{array}{ccc}
  V_1^2+V_2^2+V_3^2 & 2 V_1 V_2 & (V_1+V_2) V_4 + V_3 V_5 \\
  2 V_1 V_2 & V_1^2+V_2^2+V_3^2 & (V_1+V_2) V_4 + V_3 V_5 \\
  (V_1+V_2)V_4 + V_3 V_5 & (V_1+V_2) V_4 + V_3 V_5 & 2(V_4^2+V_5^2)
\end{array}\right].
$$
>From  the symmetry of this matrix, one eigenvalue is easy to
determine -- it corresponds to eigenvector $(1,-1,0)^T$ and equals
$$
 3U_2 -\frac{V_1^2 + V_2^2 + V_3^2- 2 V_1 V_2}{U_0}.
$$
Two other eigenvalues are also easy to obtain --they are the
solutions of some quadratic equation. Calculating $V_i$ gives
$$
V_1=2\sqrt{\frac{2}{15}}t, V_2=\sqrt{\frac{10}{3}}t,
V_3=-\sqrt{\frac{14}{5}}t, V_4=-\sqrt{\frac{14}{15}}t,
V_5=\sqrt{\frac{2}{5}}t.
$$
Energy levels are
$$
 3 U_2-4\frac{t^2}{U_0},
$$
$$
\frac12\left(9 U_2-12t^2 \pm \sqrt{144\frac{t^4}{U_0^2} + 40
\frac{t^2}{U_0} U_2 + 9U_2^2}\right).
$$

\section{Mean Field Solution for the Case of Two Bosons per Site}
\label{AppendixC}
 Our mean field state depends on $12$ variables --
$6$ complex numbers, subject to normalization
 (\ref{cnormalization}). However, energy is clearly
 the same for all states that can be transformed into each other
 by global $SU(2)$ rotations, so that gives us 3 conditions we can choose. We also have  an overall $U(1)$ phase
 freedom, so the number of independent parameters reduces to
 $12-3-1-1=7.$
We can parameterize
$$
 a_{m}=b_{m}e^{i\varphi_m}.
$$
It is convenient to choose 3 constraints to be:
$$
 b_0=0, \varphi_{-1}=\varphi_{1}.
$$
We can always find an axis of quantization for which the first
condition is satisfied and then we can satisfy the last condition
on phases by rotating along chosen axis  of quantization.
 We have 7 degrees of freedom: eight variables $ b_{\pm 2},b_{\pm 1}
,\varphi_{\pm 2},\varphi_1,\theta $ with the normalization
constraint
$$
b_2^2+b_{-2}^2+b_1^2+b_{-1}^2=1.
$$
We can express mean field energy via these degrees of freedom, and
it will be some polynomial up to the fourth order in $b_{m}$(we
won't explicitly write down this polynomial).
 First, let's make a transformation that diagonalizes the
part quadratic in $b_{\pm 2},b_{\pm 1}.$ That orthogonal
transformation is
$$
b_{-2}=\frac{-u_2+ v_2}{2}, b_{-1}=\frac{u_1- v_1}{2},
$$
$$
b_{1}=\frac{u_1 + v_1}{2},  b_{2}= \frac {u_2+v_2}{2}.
$$
Now we solve the normalization constraint by
$$
v_1 = \sin{\psi}\sin{\psi_1}, v_2= \cos{\psi} \sin{\psi_2},
$$
$$
u_1 = \sin{\psi}\cos{\psi_1}, u_2 = \cos{\psi}\cos{\psi_2}.
$$
The numerical procedure now consists of fixing the values of $t$
and $\theta$ and minimizing the expression for energy over six
angles using steepest descents method. Using this  procedure we
can numerically find energy as a function of $\theta$ for fixed
$t.$ If the minimum is attained at $\theta=0,$ then mean field
wave function is a trivial singlet. Result of this minimization
leads to the state
$$
b_m=(\frac1{\sqrt{6}},\pm \frac1{\sqrt{3}},0,\mp
\frac1{\sqrt{3}},\frac1{\sqrt{6}})^T
$$
that can be rewritten as (\ref{am}) after rotation. Though this
result was obtained numerically,  we can check analytically that
at this point all first derivatives of the energy over angles
$\psi,\psi_1$ and $\psi_2$ vanish.

\section{Large $N$ expansion}
\label{AppendixD}

In this Appendix we prove some properties of wave functions $|\psi
\rangle_N$ (\ref{Nbigwavefunction}). The normalization factor
${\cal N}$ in (\ref{Nbigwavefunction}) may be found by considering
the overlap of two states. It is sufficient to consider wave
functions constructed of only two single particle states since
rotation in the spinor state can always bring our ``pure
condensate'' wave functions to this case:
\begin{eqnarray}
|N, {\bf n}_1 \rangle &=& \frac{1}{{\cal N}} ( \cos \phi_1
a_x^{\dagger} +
\sin \phi_1 a_y^{\dagger} )^N | 0 \rangle, \nonumber\\
|N, {\bf n}_2 \rangle &=& \frac{1}{{\cal N}} ( \cos \phi_2
a_x^{\dagger} +
\sin \phi_2 a_y^{\dagger} )^N | 0 \rangle .\nonumber\\
\end{eqnarray}
We have
$$
\langle N, {\bf n}_1 | N, {\bf n}_2 \rangle = \frac{1}{{\cal N}^2}
\langle 0|  ( \cos \phi_1 a_x + \sin \phi_1 a_y )^N  ( \cos \phi_2
a_x^{\dagger} + \sin \phi_2 a_y^{\dagger} )^N | 0 \rangle =
$$
$$
\frac{1}{{\cal N}^2} \sum_{k=0}^N (C_N^k)^2 ( \cos \phi_1 \cos
\phi_2 )^k ( \sin \phi_1 \sin \phi_2)^{N-k} \langle 0 | a_x^k
a_y^{N-k} (a_x^{{\dagger}})^k (a_y^{{\dagger}})^{N-k} | 0 \rangle=
$$
$$
\frac{N!}{{\cal N}^2} \sum_{k=0}^N C_N^k ( \cos \phi_1 \cos \phi_2
)^k ( \sin \phi_1 \sin \phi_2)^{N-k} = \frac{N!}{{\cal N}^2}
cos^N(\phi_1-\phi_2) = \frac{N!}{{\cal N}^2} ({\bf n}_1 {\bf
n}_2)^{N}.
$$
Orthogonality and normalization for large $N$ now become obvious
after noting that for $({\bf n}_1 {\bf n}_2)= \cos \theta$ and
$\theta \leq \pi/2$ we have $\cos^N \theta \approx e^{-N
\theta^2/2}$.

To prove (\ref{create-annih}) we consider wave functions
\begin{eqnarray}
|N,{\bf n} \rangle &=& \frac{1}{{\cal N}_N}(n_x a_x^{{\dagger}}
+n_y a_y^{{\dagger}})^N |0 \rangle,
\nonumber\\
|N+1,{\bf n}' \rangle &=& \frac{1}{{\cal N}_{N+1}}(n'_x
a_x^{{\dagger}} +n'_y a_y^{{\dagger}})^N |0 \rangle.
\end{eqnarray}
Simple calculation gives
\begin{eqnarray}
\langle N+1, {\bf n}'| a_x^{{\dagger}}|N,{\bf n} \rangle
=\frac{(N+1)!}{{\cal N}_N {\cal N}_{N+1}}\, n'_x\, ({\bf n} {\bf
n}')^N =(N+1)^{1/2} \, n_x \, \delta_N({\bf n} - {\bf n}').
\end{eqnarray}

\end{document}